\documentclass[a4paper,11pt]{article}
\usepackage{jheppub} 
\usepackage[T1]{fontenc} 
\usepackage{amsmath,amsthm}
\usepackage{amssymb, amsfonts}
\usepackage{bbm}
\usepackage{adjustbox}
\usepackage{array}
\usepackage{graphicx}
\usepackage{dsfont}
\usepackage{relsize}
\usepackage{stmaryrd}
\usepackage{lmodern}
\usepackage{slantsc}
\usepackage{scalefnt}
\usepackage{wasysym}
\usepackage{xspace}
\usepackage{boxedminipage}
\usepackage{ifpdf}
\usepackage{multirow, hhline}
\usepackage{slashed}
\usepackage{xifthen}
\usepackage{tensor}
\usepackage{array, nicematrix}
\usepackage{tabu}
\usepackage{pbox}
\usepackage{multirow}
\usepackage{tikz}
\usepackage{overpic}
\usepackage{color}
\usepackage{diagbox}
\usepackage{makecell}
\usetikzlibrary{arrows, matrix, calc, scopes, decorations.markings}
\usepackage[mathcal]{euscript}
\usepackage{booktabs}
\usepackage{autobreak, subfig, diagbox, mathrsfs}
\usepackage{mathbbol}
\usepackage[breakable]{tcolorbox}
\tcbuselibrary{breakable}
\usepackage{hyperref}
\allowdisplaybreaks[4]

\newcommand{\ket}[1]{\left| #1 \right\rangle}

\title{Non-Abelian Particle-Loop, Fracton, and Planon Condensation in Cage-Net Models }

\author[a]{Yifei Wang}
\author[a]{Yu Zhao}
\author[b]{Yingcheng Li}
\author[c]{Hao Song\footnote{Corresponding author}}
\author[a,d]{Yidun Wan\footnote{Corresponding author}}
\affiliation[a]{State Key Laboratory of Surface Physics, Center for Astronomy and Astrophysics, Department of Physics, Center for Field Theory and Particle Physics, and Institute for Nanoelectronic devices and Quantum computing, Fudan University, Shanghai 200433, China}
\affiliation[b]{Department of Physics, National University of Singapore, 117551, Singapore}
\affiliation[c]{Institute of Theoretical Physics, Chinese Academy of Sciences, Beijing 100190, China}
\affiliation[d]{Hefei National Laboratory, Hefei 230088, China}
\emailAdd{wangyifei19@fudan.edu.cn, lyc\_phys@nus.edu.sg,  yuzhao20@fudan.edu.cn, songhao@itp.ac.cn, ydwan@fudan.edu.cn}

\abstract{
We present a framework for non-Abelian p-loop, fracton, and planon condensation in 3+1 dimensions by constructing extended cage-net fracton models using decoupled layers of the Hu-Geer-Wu (HGW) string-net model. These cage-net models extend the conventional cage-net models based on the Levin-Wen (LW) string-net model in the sense that they inherit the tail degrees of freedom of the HGW models, which are essential for completely describing the internal spaces of quasiparticles. This approach allows us to explicitly derive the quasiparticle spectra of the cage-net models by projecting those of the parent 2D HGW layers. Utilizing this framework, we can condense the p-loops formed by non-Abelian anyons within a fracton phase. Specifically, we construct the condensation projector for $(\sigma\bar{\sigma}, 1)$-loops within the extended Ising Cage-Net (ICN) model. We demonstrate that condensing these non-Abelian loops drives a phase transition that maps the ICN model to the X-cube (XC) model defined on a truncated cubic lattice, a process that explicitly reveals the splitting of non-Abelian planons into distinct sub-dimensional excitations. Furthermore, our framework extends to the condensation of fractons and planons: we demonstrate that in the ICN model fracton condensation drives the decoupling of the 3D fracton order back into isolated 2D topological order layers, while planon condensation collapses the system entirely into a trivial phase. Our results establish a concrete Hamiltonian mechanism for phase transitions between distinct fracton orders and provide a generalizable method for analyzing the evolution of sub-dimensional excitations.
}

\begin{document}
\maketitle
\flushbottom

\section{Introduction}

Fracton orders (FOs) are exotic gapped quantum phases of matter in $3+1$ dimensions and higher that greatly expand the concept of topological orders \cite{vijay_fracton_2016, haah_local_2011, nandkishore_fractons_2019, pretko_fracton_2020}. They have stimulated active research across quantum information and quantum memory \cite{Bravyi2013,yoshida_exotic_2013,Brown2020,song_optimal_2022,Canossa2026}, quantum field theory \cite{Pretko2017,Pretko2017_2,Seiberg2020,seiberg_exotic_2021,seiberg_exotic_2021-1,gorantla_more_2020,ma_fracton_2018,bulmash_higgs_2018,you_fractonic_2020,slagle_quantum_2017,slagle_foliated_2019,slagle2021,Ma_2022}, and many-body physics \cite{chamon_quantum_2005,vijay_new_2015,slagle_fracton_2017,ma_fracton_2017,Pretko2018,prem_cage-net_2019,song_fracton_2024,canossa_exotic_2024,Xu2022,prem_pinch_2018,Yan2020,li_fracton_2020,li_fracton_2021,Chen_Ye_2020,Chen_Ye_2021,zhu2023, gromov_multipole_2019, gromov_fracton_2020,zerba_emergent_2025}.
Among various approaches \cite{haah_commuting_2013,haah_bifurcation_2014,shirley_fracton_2018,williamson_fractal_2016,song_twisted_2019,shirley_foliated_2019,aasen_topological_2020,wen_systematic_2020,wang_nonliquid_2022,song_topological_2023,Vadali2024,tantivasadakarn_hybrid_2021,tantivasadakarn_non-abelian_2021} to fracton orders, p-loop condensation provides a useful route to constructing and understanding many representative models \cite{ma_fracton_2017,prem_cage-net_2019,vijay2017isotropiclayerconstructionphase}.
Given a system comprising three mutually orthogonal and penetrating stacks of decoupled layers of topological orders, a p-loop\footnote{It is usually called a p-string in the literature, but it is in fact a closed loop, so in this paper, we shall call it a p-loop.} is a 1-dimensional loop formed by specific anyons residing in two or more intersecting layers. Condensing certain types of p-loops drives the system through a phase transition into a certain FO. Models of FOs constructed via Abelian p-loop condensation include the X-cube (XC) model and the Ising cage-net (ICN) model. Nevertheless, two interesting questions remain open.  First, the p-loops having been considered so far consist of Abelian anyons only, so is it possible to condense p-loops formed by non-Abelian anyons? Second, the phase-transitions caused by p-loop condensations thus far considered drive decoupled systems of TOs to FOs, and phase transitions between two nontrivial FOs have not been studied in the literature, so how can one realize a phase transition between two FOs?

In this paper, we tackle these challenges in one scoop: by replacing the traditional LW model with the HGW model, our construction inherits tail degrees of freedom that explicitly resolve the complete dyonic spectrum (internal gauge spaces) of the underlying anyons.  Within this extended cage-net model, we can rigorously construct the non-Abelian p-loop condensation operator by explicitly identifying the internal dyonic sectors of the non-Abelian sub-dimensional quasiparticles and selectively extracting the flux-free sectors. Although our theoretical construction is generic, to be specific and for showcasing, we consider the ICN model as the parent model and perform non-Abelian p-loop condensation. By condensing $(\sigma\bar{\sigma}, 1)$-loops, the parent ICN model undergoes a phase transition into a child FO model, which we prove to be the exact XC model defined on a truncated cubic lattice. 

This transition allows us to establish a precise relationship between the sub-dimensional quasiparticles of the parent and child FOs. Crucially, we show that non-Abelian planons in the ICN model undergo splitting into distinct sub-dimensional excitations (such as strictly immobile fractons) in the XC model, alongside phenomena such as identification and confinement. These mechanisms serve as the 3D fractonic counterparts to the phenomena arising in 2D anyon condensation. Furthermore, we expand our generic framework to complete the hierarchy of sub-dimensional condensations: we demonstrate that the direct condensation of fractons drives an decoupling from the 3D ICN model back into isolated 2D topological order model layers, while the condensation of non-Abelian planons collapses the system entirely into a trivial phase. Together, these results provide a comprehensive Hamiltonian roadmap for the dynamic manipulation of spatial dimensions and mobility constraints in topological order and fracton models.

\section{Background and Our Approach}
Models of fracton order are often divided into two types. ``Type-I'', such as the XC model and checkerboard model \cite{vijay_fracton_2016}, can support fractons, lineons and planons, while ``Type-II'', such as Haah's code \cite{vijay_fracton_2016,haah_local_2011}, can only support fractons. Recent studies show that type-I fracton models like the XC model and ICN model \cite{prem_cage-net_2019,ma_ground_2023,wang_renormalization_2023} can be constructed from decoupled layers of $2+1$d models that describe topological order (TO) via a procedure known as ``p-loop condensation'' \cite{ma_fracton_2017,vijay2017isotropiclayerconstructionphase,prem_cage-net_2019,zhu2023}. Consider lattice model for example. Given a parent model (consisting decoupled TO model layers), one can add to its Hamiltonian a coupling term that acts on every edge where two perpendicular layers are intersected (often termed as a principal edge). This coupling term can be written as the product of certain Abelian p-loop creation operators $P_E$ on principal edges $E$. The coupling term is designed to be a projector, ensuring that at infinite coupling strength, the new ground states are the $+1$ eigenstates of $P_E$ and are thus coherent states with arbitrarily many condensed p-loops in the lattice. The resultant child model after p-loop condensation is a fracton model. The Hibert space and Hamiltonian of the child fracton model can be obtained by applying the projector to those of the parent model. While p-loop condensation not only shows it is possible to construct ${3+1}$d fracton models from ${2+1}$d TO models, it also allows us to study fracton models by using known data of the parent ${2+1}$d TO model layers.

Nonetheless, formulating the condensation of p-loops comprising non-Abelian anyons is difficult. Unlike Abelian p-loops, which can fully condense, non-Abelian p-loops in general cannot fully condense but have non-condensable parts because of their complex fusion rules. This complexity makes it difficult to formulate a non-Abelian p-loop condensation operator capable of generating coherent states. Besides, because non-Abelian anyons have dyonic sectors (charge-flux composites), and the condensation of charged p-loops remains poorly understood. Take the doubled Ising string-net model as an example, its non-Abelian anyon $\sigma \Bar{\sigma}$ has a dyonic sector $(\sigma\bar\sigma,1)$ carrying trivial charge 1 and one $(\sigma\bar\sigma,\psi)$ carrying nontrivial charge $\psi$. Two $\sigma \Bar{\sigma}$ fuse as
\begin{equation*}
\sigma \Bar{\sigma } \times \sigma \Bar{\sigma } =1\Bar{1} +1\Bar{\psi } +\psi \Bar{1} +\psi \Bar{\psi } .
\end{equation*}
If one wants to condense $\sigma\bar\sigma$, one would expect that all its fusion products should be condensed too; however, $1\bar\psi$ and $\psi\bar 1$ are non-condensable, as they are fermions.

In this paper, we shall formulate a general theory of p-loop condensation applicable to both Abelian and non-Abelian cases. We focus on cage-net models \cite{prem_cage-net_2019}, as they are constructed from the LW models -- the most general model for describing doubled topological orders in $2+1$ dimensions. Formulating such a theory requires complete knowledge of the \textit{full dyon spectrum} of the string-net model. This would need an extended version of the LW model, known as the Hu-Geer-Wu (HGW) model\cite{hu_full_2018,zhao_landau-ginzburg_2025}, which adds to each string-net plaquette a tail degree of freedom, enabling explicit description of dyonic sectors. Condensation of non-Abelian anyons with explicit handling of dynonic sectors in the HGW model has been studied by Zhao et al\cite{zhao_characteristic_2023,zhao_nonabelian_2025}.

Despite being general, we shall showcase our approach by a concrete example: We shall begin with decoupled layers of the doubled Ising HGW model, then condense the p-loops where $\text{p}=\psi\bar\psi$ to obtain a cage-net model, and finally condense the non-Ablelian p-loop where $\text{p}=(\sigma\bar\sigma,1)$ in this resultant cage-net model.  Condensing $\psi \Bar{\psi}$-loops in decoupled layers of the doubled Ising HGW model will lead to the extended ICN model, which has a tail degree of freedom inherit from the HGW model. We then try to condense the $(\sigma \Bar{\sigma},1)$-loop in the extended cage-net model. This could not be done in the usual cage-net model because one does not know how to express the $(\sigma \Bar{\sigma} ,1)$-loop creation operator. In our resultant extended cage-net model however, we find that the $(\sigma \Bar{\sigma} ,1)$-loop creation operator in the decoupled layers of the doubled-Ising HGW model survives the $\psi \Bar{\psi}$-loop condensation and thus retains its form in the extended cage-net model. So, we can then construct the $(\sigma \Bar{\sigma } ,1)$-loop condensation term using the $(\sigma \Bar{\sigma} ,1)$-loop creation operator and add the term with a coupling strength to the extended cage-net Hamiltonian. We design the $(\sigma \Bar{\sigma } ,1)$-loop condensation term to make sure that at infinite coupling strength, the new ground states are coherent states with arbitrarily many condensed $(\sigma \Bar{\sigma } ,1)$-loops. We find that the $(\sigma \Bar{\sigma },1)$-loop condensation triggers a fracton phase transition from the extended ICN model to the XC model defined on the truncated cubic lattice.

\section{Anyon condensation in the HGW string-net model}
\label{sec: anyon cond HGW}
In this section, we briefly review the HGW string-net model introduced in \cite{hu_full_2018} and how to do anyon condensation in the doubled Ising HGW string-net model\cite{zhao_characteristic_2023}. We will also briefly review the phenomena in phase transition between the doubled Ising string-net model and the $\mathbb{Z}_2$ toric code triggered by $\psi\Bar{\psi}$ anyon condensation.
\subsection{HGW string-net model}
\begin{figure}
    \centering
    \includegraphics[width=1\linewidth]{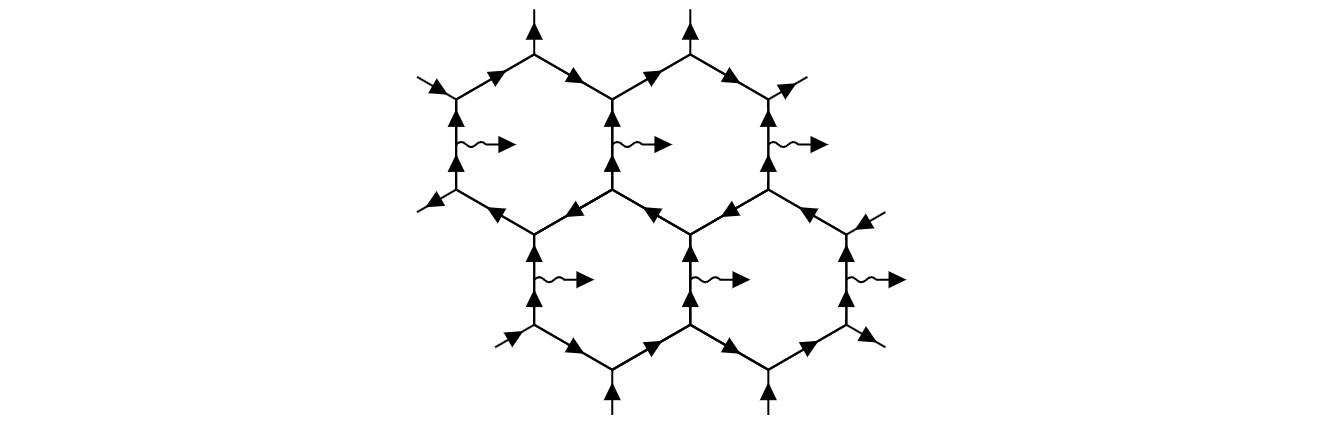}
    \caption{Part of the HGW model lattice. A tail (wavy line) is attached to an arbitrary edge of every plaquette.}
    \label{fig:HGW}
\end{figure}

The HGW string-net model is defined on a 2-dimensional oriented trivalent lattice. A tail is assigned to every plaquette and attached to any of the plaquette's edges, pointing inward, see Figure \ref{fig:HGW}. The different selections of the edge to which the tail is attached are equivalent, and tails (sometimes we may encounter multiple tails within a single plaquette in auxiliary states) within one plaquette can be contracted into one tail(see Appendix\ref{HGW theory}). Each edge or tail carries a label which takes value in the set $L_\mathcal{F}$, consisting simple objects of the model's input UFC $\mathcal{F}$. The Hilbert space is spanned by all possible assignments of the labels, while under the constraint that fusion rules are always satisfied at every vertex. In the doubled Ising (DI for short) case, the labels take value in the set $L_\text{DI}=\{1,\psi,\sigma\}$. The Hilbert space $\mathcal{H}_\text{DI}$ is spanned by all possible assignments of $1$, $\psi$ and $\sigma$ to all edges and tails, with the fusion rules:
\begin{equation}
    \delta_{\sigma \sigma \sigma}=\delta_{1 \psi \psi}=\delta_{1 \sigma \sigma}=\delta_{\psi \sigma \sigma}=1,
    \label{ising fusion rule}
\end{equation}
satisfied at every vertex.

The Hamiltonian reads:
\begin{equation}
    H_\text{DI}:=-\sum_{\text {Plaquettes } P} Q_P,
\end{equation}
where the operators $Q_P$ are detailed in Appendix \ref{HGW theory}. 
These operators are commuting projectors so the Hamiltonian is exactly solvable. The ground states $\ket{\Phi}$ are common $+1$ eigenstates of all $Q_P$ operators. An excited state $\ket{\varphi}$ is another common eigenstate that satisfies $Q_P\ket{\varphi}=0$ for one or more plaquettes $P$, each of which bears an anyon. Ground state is also said to have a trivial anyon in each plaquette. There are nine anyon species in the doubled Ising topological order:

\begin{equation}
    1\Bar{1}, 1\Bar{\psi}, 1\Bar{\sigma}, \psi\Bar{1}, \psi\Bar{\sigma}, \sigma\Bar{1}, \sigma\Bar{\psi}, \psi\Bar{\psi}, \sigma\Bar{\sigma},
\end{equation}
where $1\Bar{1}$ is the trivial anyon.

Moreover, we need to introduce the concept of anyon's \textit{internal space}. In the HGW string-net model, an excited state is determined by the anyon species (labeled in each plaquette) and the anyon's internal charge (labeled by the d.o.f on the tail where the anyon resides). A non-Abelian anyon carries more than one charge type and is thus represented on a certain multi-dimensional Hilbert subspace of excited states of the model. In other words the HGW string-net model represent an anyon $J$ as a $\textit{dyon}$, a pair $(J,p)$ comprising the anyon's species $J$ and its internal charge $p\in L$. The doubled Ising string-net model has ten dyon types:
\begin{equation}
    (1\Bar{1},1), (1\Bar{\psi},\psi), (1\Bar{\sigma},\sigma), (\psi\Bar{1},\psi), (\psi\Bar{\sigma},\sigma), (\sigma\Bar{1},\sigma), (\sigma\Bar{\psi},\sigma), (\psi\Bar{\psi},1), (\sigma\Bar{\sigma},1), (\sigma\Bar{\sigma},\psi),
\end{equation}
where the anyon $\sigma\Bar{\sigma}$ can have a charge of either $1$ or $\psi$ in the model.

The $\textit{elementary excitation state}$ (EES for short) is the excited state with at most two dyons. In an EES all tails take trivial value $1$ except those where the two dyons reside, thus we can omit the tails irrelevant to these dyons. Each EES $\ket{\phi}$ can be obtained by acting a $\textit{ribbon operator}$ $W_L$ on the ground state $\ket{\Phi}$:
\begin{equation}
    \ket{\varphi}=W_L\ket{\Phi}.
\end{equation}
The ribbon operator $W_L$ is defined along a path $L$, which cross one or more edges in the lattice, and creates a pair of dyons at the two ends of $L$. Since we can always concatenate shorter ribbon operators to a longer one (see Appendix~\ref{HGW theory}), it suffices to study the shortest ribbon operator $W_E^{J;p,q}$ (also dubbed as creation operator), which crosses only one edge $E$ and creates a pair of dyons $(J,p)$ and $(J,q)$ in the two adjacent plaquettes. In the doubled Ising HGW model, we write the creation operator around edge $E$ and its corresponding EES as
\begin{equation}
    \ket{J_\text{DI};p,q}_E=\vcenter{\hbox{\includegraphics[height=10ex]{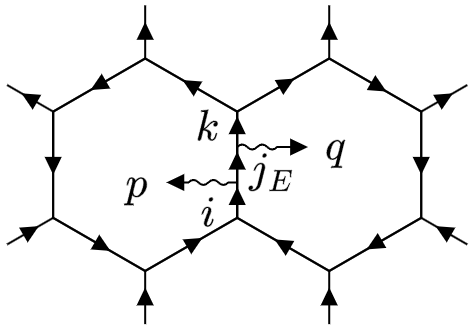}}}:=W_E^{J_\text{DI};p,q}\ket{\Phi}_\text{DI}.
    \label{HGW EES}
\end{equation}
The matrix elements of $W_E^{J_\text{DI};p,q}$ are detailed in Appendix \ref{doubled ising data}. 

There are twelve creation operators in the doubled Ising phase in total: crossing any edge $E$, for each $J_\text{DI}\neq\sigma\Bar{\sigma}$, there is only one such creation operator $W_E^{J_\text{DI};p,q}$ with $p=q$; for each $J_\text{DI}=\sigma\Bar{\sigma}$, the corresponding $p$ and $q$ can both take values in the set $\{1,\psi\}$, resulting four different such creation operators $W_E^{\sigma\Bar{\sigma};p,q}$ with $p,q\in\{1,\psi\}$. The four EESs $\ket{\sigma\Bar{\sigma};p,q}_\text{DI}=W_E^{\sigma\Bar{\sigma};p,q}\ket{\Phi}_\text{DI}$ with $p,q\in\{1,\psi\}$ are degenerate \cite{zhao_characteristic_2023}, while other EESs are non-degenerate. 

\subsection{Anyon condensation}
Anyon condensation is the main mechanism to trigger topological phase transition. Physically, after anyon condensation, the new vacuum states of the child model become coherent states containing arbitrarily many condensed anyons from the parent model. To realize such coherent states, we can add an anyon-condensation term to the parent Hamiltonian $H_\text{parent}$:

\begin{equation}
    H_{\text{parent}} \rightarrow H_{\text{parent}}-\lim _{\Lambda \rightarrow \infty} \Lambda \sum_{\text {Edges } E} P_E,
    \label{3.7}
\end{equation}
\begin{equation}
    P_E:=\sum_{\text {Condensed } J} \quad \sum_{ J^{\prime}\text{s charges } p, q} \frac{\pi_J^{p q}}{d_p d_q} W_E^{J ; p,q} .
    \label{anyon cond term}
\end{equation}
Here $d_p$ ($d_q$) is the quantum dimension of the charge $p$ ($q$). The coefficients $\pi_J^{pq}$ are chosen to make $P_E$ a projector. For $\Lambda\rightarrow\infty$, the new ground states become coherent states with arbitrarily many condensed anyons $J$ throughout the lattice. The child model is also described by a string-net model, with the child Hilbert space $\mathcal{H}_\text{child}$ and the child Hamiltonian $H_\text{child}$ given by:
\begin{equation}
\mathcal{H}_{\text{child }}= P \mathcal{H}_{\text{parent}},\quad H_{\text{child }}=P H_{\text{parent}}P,\quad P:=\left[\prod_{\text{Edges } E} P_E\right] .
\end{equation}
The spectrum of the child model is also determined by this projection. In other words, in the child model, the ground states $\ket{\Phi}_\text{child}$, the creation operators $W^{J_{\text{child}};p_{\text{child}},q_{\text{child}}}$, and the corresponding EESs $\ket{J_\text{child};p_{\text{child}},q_{\text{child}}}$ can be obtained by applying projector $P$ to those in the parent model.

Consider the case of $\textit{chargeon condensation}$. Since a chargeon has trivial charge $1\in L_\mathcal{F}$, each basis state $\ket{\phi}$ of the string-net model is an eigenstate of the chargeon creation operators $W_E^{J;1,1}$:
\begin{equation}
    W_E^{J;1,1}\ket{\phi}=w_J (j_E)\ket{\phi}, \quad \ket{\phi}=\vcenter{\hbox{\includegraphics[height=10ex]{EES_HGW.png}}}
\end{equation}
where $j_E$ is the d.o.f on edge $E$, and $w_J (j_E)\in \mathbb{C}$ is the eigenvalue of $W_E^{J;1,1}$. 
To do chargeon condensation, we first choose a subset $L_\mathcal{S} \subset L_\mathcal{F}$ of simple objects that are closed under fusion. This subset $L_\mathcal{S}$ generates a full subcategory $\mathcal{S}$ of the parent input UFC $\mathcal{F}$, and $\mathcal{S}$ can serve as input UFC of the child model, whose basic d.o.fs take values from the simple objects in the subset $L_\mathcal{S}$. The condensation projector $P_E$ in \eqref{anyon cond term} then becomes:
\begin{equation}
    P_E^{\mathcal{S} \mid \mathcal{F}}\ket{\phi}=\left[\sum_{\text {Condensed chargeons } J} \pi_J^{11} w_J\left(j_E\right)\right]\ket{\phi}=\delta_{j_E \in L_{\mathcal{S}}}\ket{\phi}
    \label{anyon cond eq}
\end{equation}
where $\delta$ is the Kronecker symbol. Equation \eqref{anyon cond eq} always has a unique solution because the number of chargeon species in a string-net model always equals the number of simple objects\cite{hu_full_2018}.

\paragraph{Phenomena due to anyon condensation}
Several phenomena arise due to anyon condensation \cite{bais_condensate-induced_2009,eliens_diagrammatics_2014,hu_anyon_2022,gu_unified_2014,zhao_nonabelian_2025}:

1. \textit{Splitting}: Certain anyons (include those condensed) may split into multiple sectors and become different anyon species in the child order during anyon condensation.

2. \textit{Identification}: As condensed sectors become the new vacuum, two or more types of sectors related by fusing with a condensed sector in the parent order can no longer be distinguished in the child order and thus should be identified as the same type of child anyon.

3. \textit{Confinement}: Anyons braiding nontrivially with the condensate become confined in the child order because the new vacuum should not be disturbed by moving anyons around.

For example, Table \ref{tab:anyon cond pheno} records the phenomena mentioned above when condensing $\psi\Bar{\psi}$ in the doubled Ising topological order \cite{zhao_characteristic_2023,hu_full_2018}.

\begin{table}[!h]
    \centering
    
    \begin{tabular}{|c|c|c|c|c|}
        \hline 
        & \multicolumn{2}{c|}{Identification} & Splitting & Confinement \\ 
        \hline 
        $J_{\text{DI}}$ & $1\Bar{1}, \psi \Bar{\psi}$ & $1\Bar{\psi}, \psi \Bar{1}$ & $\sigma \Bar{\sigma}$ & $1\Bar{\sigma}, \psi \Bar{\sigma}, \sigma \Bar{1}, \sigma \Bar{\psi}$ \\ 
        \hline 
        $J_{\text{TC}}$ & $1$ & $\epsilon$ & $e,m$ & $\emptyset$ \\ 
        \hline
    \end{tabular}
    \caption{Summary of topological occurring during the phase transition from doubled Ising phase to $\mathbb{Z}_2$ toric code driven by $\psi\bar{\psi}$ condensation. The first row ($J_{\text{DI}}$) lists the relevant anyon types in the parent phase, while the second row ($J_{\text{TC}}$) indicates their corresponding anyon types in the child phase. The columns categorize the transitions into three mechanisms: Identification, where fusion with the condensate renders distinct parent sectors indistinguishable (e.g., both $1\bar{1}$ and $\psi\bar{\psi}$ map to the vacuum $1$); Splitting, where a single parent anyon ($\sigma\bar{\sigma}$) decomposes into distinct anyons ($e$ and $m$) in the child phase; and Confinement, where sectors braiding non-trivially with the condensate (such as those containing a single $\sigma$ charge) are confined and removed from the low-energy spectrum ($\emptyset$).}\label{tab:anyon cond pheno}
\end{table}

\section{Extended ICN model from the doubled-Ising HGW string-net model}

A cage-net fracton model describes a 3-dimensional gapped fracton phase that hosts not only immobile fracton excitations but also non-Abelian particles with restricted mobility. An example of this model -- the ICN model -- was constructed from decoupled layers of the 2-dimensional doubled-Ising LW model \cite{prem_cage-net_2019,ma_ground_2023,wang_renormalization_2023} via a process known as p-loop condensation, which condenses loops consisting of $\psi\bar\psi$ anyons across the layers. Unfortunately, this construction lacks a correct condensation procedure at the Hamiltonian level. The reason for this incorrectness is that the LW model is not suitable for performing anyon condensation. As reviewed in Section \ref{sec: anyon cond HGW}, the HGW model is the right model to construct anyon-condensation terms at the Hamiltonian level. 

In this section, we shall reconstruct the ICN model\footnote{General cage-net models can be constructed similarly, but we focus on the Ising case in this paper.} from decoupled layers of the HGW model with the input Ising UFC, by condensing $\psi\Bar{\psi}$-loops throughout the 3-dimensional lattice. In contrast to the cage-net models in the literature, in the cage-net models constructed in our way, we will be able to perform nonabelian p-loop condensation and study the resultant fractonic phase transition. We shall term our cage-net models as extended cage-net models.   

\subsection{Decoupled layers of HGW string-net model and p-loop condensation operator}

To obtain a 3-dimensional model consisting of decoupled layers of the HGW model (dHGW for short) with input UFC $\mathcal{F}$, we need to slightly modify the original HGW model defined on the honeycomb lattice, such that it is properly defined on the 2-dimensional truncated square lattice. Three modifications are needed:
\begin{enumerate}
\item A tail is assigned to every octagonal and square plaquette, see Figure \ref{fig:trunc square}.

\begin{figure}
    \centering
    \includegraphics[width=1\linewidth]{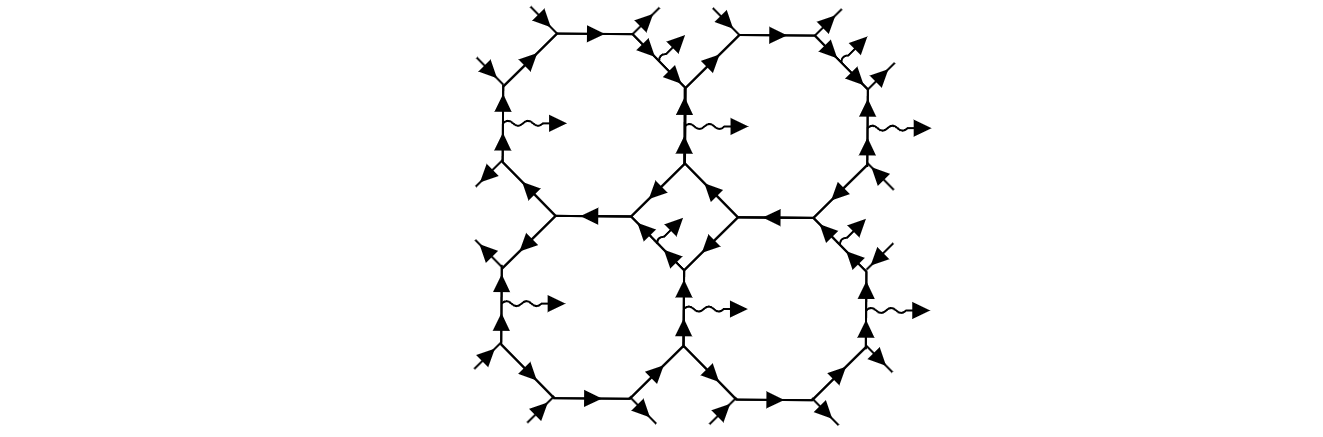}
    \caption{Configuration of the HGW model on the $2$-dimensional truncated square lattice. The lattice geometry is constructed from trivalent vertices and features two distinct types of plaquettes: large octagonal plaquettes and smaller square (or diamond-like) plaquettes. A tail degree of freedom, depicted as a wavy line, is attached to every plaquette of both types and points inward, satisfying the requirements for the HGW formalism in this modified geometry.}
    \label{fig:trunc square}
\end{figure}

\item The Hamiltonian now takes the form:
\begin{equation}
    H_{2\text{d HGW}}:=-\sum_{\text{Octagonal Plaquettes}\  P_o}Q_{P_o}-\sum_{\text{Diamond Plaquettes}\   P_d}Q_{P_d},
    \label{HGW trunc}
\end{equation}
where operators $Q_{P_o}$ and $Q_{P_d}$ are detailed in Appendix \ref{HGW theory}.
\item The ground states $\ket{\Phi}$ are the common $+1$ eigenstates of all $Q_{P_o}$ and $Q_{P_d}$ operators.
%
\end{enumerate}

\begin{figure}
    \centering
    \includegraphics[width=\linewidth]{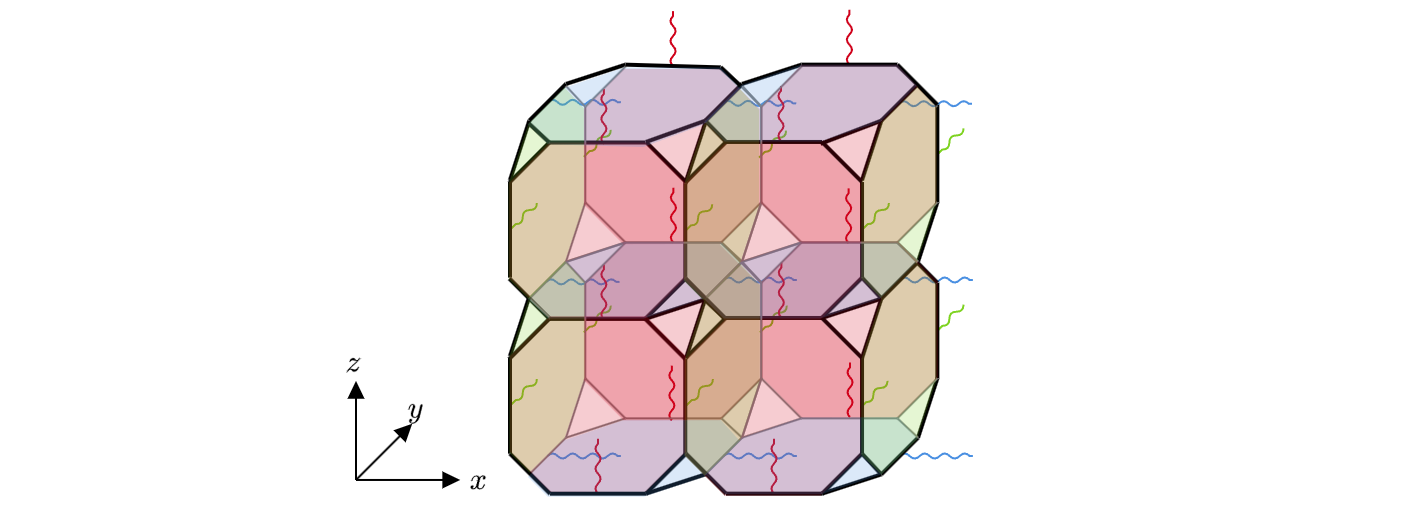}
    \caption{$3$D truncated cubic lattice formed by intersecting three orthogonal stacks of $2$D HGW layers. The layers are color-coded according to their orientation: green, red, and blue shading indicates layers normal to the $x$, $y$, and $z$ directions, respectively. The lattice structure features principal edges which are formed at the intersection of two perpendicular layers and carry a pair of labels ($j_E,k_E \in \mathcal{F}$). A tail (wavy line) is specifically attached to each principal edge and directed into the adjacent octagonal plaquette. This configuration ensures that every octagonal plaquette possesses exactly one tail , essential for defining the dyonic sectors in the extended cage-net model.}
    \label{fig:trunc cubic}
\end{figure}

Then we stack such layers along each of the three perpendicular directions $\mu \in \{x,y,z\}$ with a fixed distance. We then join the three perpendicular stacks to form a truncated cubic lattice, see Figure \ref{fig:trunc cubic}.

The edges that are along the $x$, $y$, or $z$ directions are called \textit{principal edges}. Each principal edge $E$ in the 3-dimensional truncated cubic lattice consists of two edges respectively belonging to two intersecting 2-dimensional truncated square lattices, so it carries two labels $j_E, k_E\in \mathcal{F}\times \mathcal{F}$. One tail is attached to every principal edge, ensuring that every octagonal plaquette has exactly one tail (shown in Figure \ref{fig:trunc cubic}). If two tails attach to the same principal edge and are perpendicular to each other, they must be anchored at the exact same position on that edge. In contrast to principal edges, any tail or non-principal edge can only lie in exactly one plaquette and thus carries a single label taking value in the set $\mathcal{F}$.
We label each plane as $(E,\mu)$, where $\mu=x,y,z$ is the normal direction of the plane containing the principle edge $E$. The Hilbert space $\mathcal{H}_\text{dHGW}$ is spanned by all possible assignments of $L_\text{dHGW}$ to all principal edges and $L_{2\text{d HGW}}$ to all tails and other edges, with the fusion rules satisfied at every vertex.
The Hamiltonian of the decoupld HGW is simply the sum of \eqref{HGW trunc} over all planes:
\begin{equation}
    H_\text{dHGW}=\sum_{(E,\mu)}{H^{(E,\mu)}_{2\text{d HGW}}}.
\end{equation}
An excited state in the dHGW around a principal edge $E$ pointing towards $x$ direction can be written as:
\begin{equation}
    \vcenter{\hbox{\includegraphics[height=16ex]{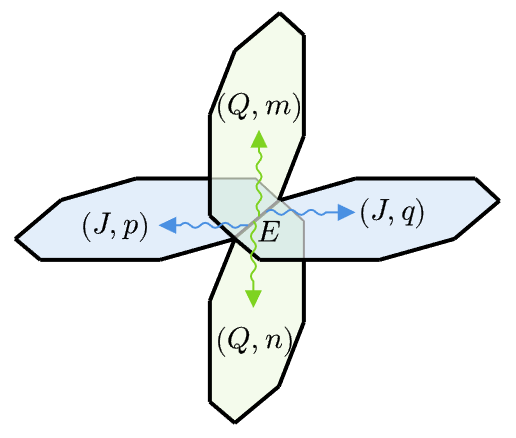}}}:=W_E^{J;p,q}\ket{\Phi}_{(E,z)}V_E^{Q;m,n}\ket{\Phi}_{(E,y)}
\end{equation}
where $W_E^{J;p,q}$ and $V_E^{Q;m,n}$ are creation operators defined in \eqref{HGW EES}, living in the two perpendicular planes that shares one principal edge $E$. Ground states $\ket{\Phi}_{(E,z)}$ and $\ket{\Phi}_{(E,y)}$ are associated with the HGW models on the planes $(E,z)$ and $(E,y)$. Excited states around principal edge $E$ pointing towards $y$ direction or $z$ direction can be written in a similar way. In the case of chargeons we have $p=q=1$, and we may write $W^{J;1,1}_E$ as $W^J_E$ for convenience. 

\begin{figure}[!ht]
    \centering   \includegraphics[width=0.85\linewidth]{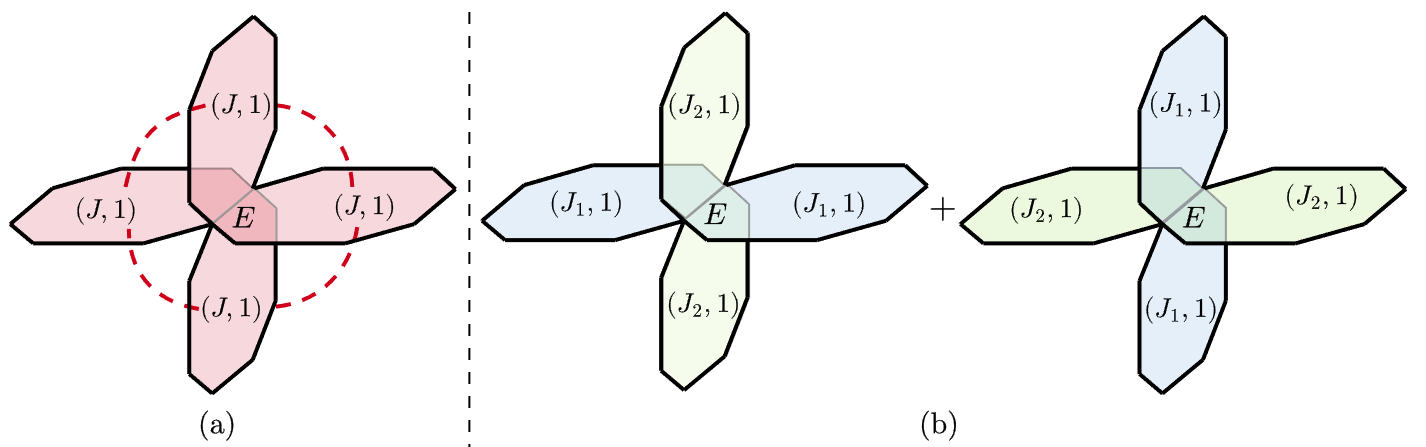}
    \caption{Graphical representation of the local terms comprising the p-loop condensation projector $P_E$ acting on a principal edge $E$. \textbf{(a)} A $J$-loop generated by the operator $W_{E}^{J} V_{E}^{J}$, creating a loop of four $J$-chargeons around the principal edge. \textbf{(b)} A $J_1-J_2$ --- cross generated by the operator term $W_{E}^{J_1} V_{E}^{J_2} +W_{E}^{J_2} V_{E}^{J_1}$. This configuration represents a symmetric superposition of two states where chargeons $J_1$ and $J_2$ are exchanged across the intersecting planes, ensuring the projector preserves cubic symmetry.}
    \label{fig:J loop cross}
\end{figure}

To trigger p-loop condensation in the dHGW, we add a \textit{global p-loop condensation term} to the parent Hamiltonian:
\begin{equation}
    H_\text{dHGW}\rightarrow H_\text{dHGW}-\lim_{\Lambda\rightarrow\infty}\Lambda P, \quad P=\prod_{\text{Principal Edges E}}{P_E},
    \label{dHGW Hamil with p-loop}
\end{equation}
where the global p-loop condensation term $P$ is a sum of all local projectors $P_E$ defined on every principal edge $E$. This global p-loop condensation term would couple all the $2$d HGW string-net model layers and turn them into a $3$d model. In this paper, we shall restrict ourselves to the case of pure chargeon condensation only, i.e., $p,q,m,n\equiv 1$. In other words, $P_E$ generates p-loops around principal edge $E$:

\begin{equation}
    P_{E} := \underbrace{\sum\limits _{\{J\}} \pi _{JJ} W_{E}^{J} V_{E}^{J}}_{P_E^\text{loop}} \ +\underbrace{\sum\limits _{J_1,J_2\in \{J\}, \ J_1\neq J_2} \pi _{J_1 J_2}\left( W_{E}^{J_1} V_{E}^{J_2} +W_{E}^{J_2} V_{E}^{J_1}\right)}_{P_E^\text{cross}}.
    \label{eq:cLoopPE}
\end{equation}
where $\{J\}$ is the set of chargeons to be condensed. Equation \eqref{eq:cLoopPE} promotes \eqref{anyon cond term} to the case of 3 dimensions, while restricted to chargeons. Operator $W_{E}^{J} V_{E}^{J}$ in $P_E^\text{loop}$ creates a p-loop consisting of four $J$-chargeons around principal edge $E$, as in Fig. \ref{fig:J loop cross}(a). We call this p-loop a $J$-loop. Operator $W_{E}^{J_1} V_{E}^{J_2} +W_{E}^{J_2} V_{E}^{J_1}$ in $P_E^\text{cross}$ creates what we call a $J_1-J_2$-cross, as depicted in Fig. \ref{fig:J loop cross}(b). Note that a $J_1-J_2$-cross is a superposition of two states, symmetric under the exchange of $J_1$ and $J_2$ to preserve the cubic symmetry. The coefficients $\pi_{JJ},\pi_{J_1 J_2} \in \mathbb{C}$ in \eqref{eq:cLoopPE} make $P_E$ a projector. 

\begin{figure}[!ht]
    \centering   \includegraphics[width=1\linewidth]{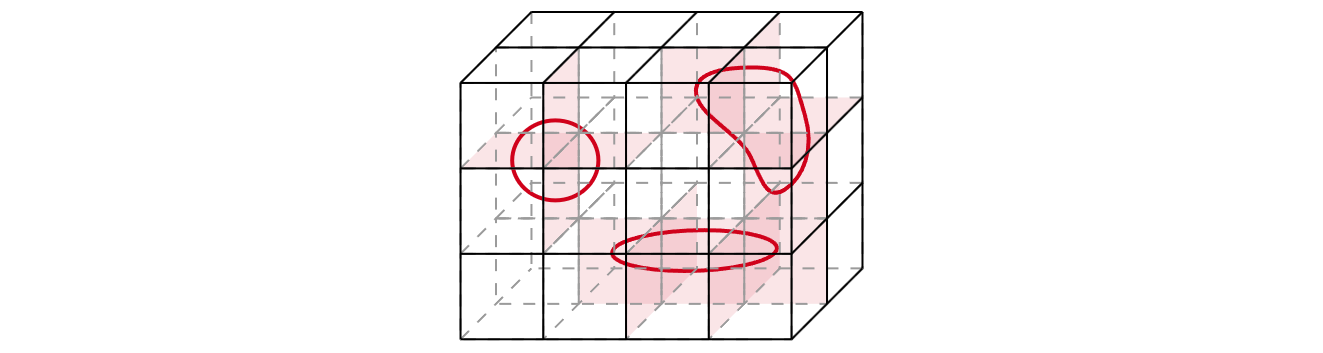}
    \caption{Illustration of certain p-loop configurations. The p-loop condensate is a superposition of various p-loop configurations fluctuating throughout the $3$D lattice. Thick red loops represent the condensed p-loops formed along the principal edges, while the light red plaquettes indicate the underlying chargeon excitations that constitute these loops.}
    \label{fig:p-loop cond}
\end{figure}

We observe that all terms in $P_E$ commute. In Abelian cases, when two p-loops fuse into a longer p-loop, there will be no extra excitations enclosed by the longer p-loop, and the coefficients of $J_1-J_2$ cross terms are zero. In non-Abelian cases, the $J_1-J_2$ cross terms in $P_E$ will cancel out any unwanted nontrivial excitations in fusing two p-loops. When $\Lambda$ goes to infinity, the system's ground states are dominated by the common $+1$ eigenstates of all $P_E$ projectors. Such a ground state is a superposition of all possible configurations (see figure~\ref{fig:p-loop cond} for an example) of all kinds of p-loops, namely a p-loop condensate.

Our goal in this section is to turn the dHGW model into a extended cage-net model by p-loop condensation. This is realized if the projector $P_E$ \eqref{eq:cLoopPE} acts on a principal edge $E$ in a way such that 
\begin{equation}
    P_E \ket{\psi}_E\ket{\varphi}_E =\delta_{j_E| k_E \in L_{\text{CN}}}\ket{\psi}_E\ket{\varphi}_E.
    \label{p-loop proj}
\end{equation}
That is, $P_E$ projects the degree of freedom on each principal edge of the dHGW model to that of a CN model. This is possible because $P_E$ \eqref{eq:cLoopPE} comprises only chargeon creation operators $W^J_E$ and $V^J_E$, of which any principal edge state $\ket{\psi}\ket{\varphi}$ is an eigenstate; hence, 
\begin{equation}
\begin{split}
    P_E \ket{\psi}_E\ket{\varphi}_E
    = \bigg[ & \sum_J \pi_{JJ} w_J \left(j_E\right) v_J \left(k_E\right) \\
    & + \sum_{J\neq Q} \pi_{JQ} \left( w_J \left(j_E\right) v_Q \left(k_E\right) + w_Q \left(j_E\right) v_J \left(k_E\right) \right) \bigg] \ket{\psi}_E\ket{\varphi}_E,
\end{split}
\end{equation}
where $w_J(j_E)$ and $v_J(j_E)$ are the eigenvalues of $W^J_E$ and $V^J_E$. It then suffices to find appropriate coefficients $\pi_{J J}$ and $\pi_{J Q}$ to fulfill \eqref{p-loop proj}.

In what follows, we shall consider two specific cases: 1) constructing the XC model from the decoupled $\mathbb{Z}_2$ HGW model by condensing the $e$-loops, and 2) constructing the ICN model from the decoupled Ising HGW model by condensing the $\psi\bar\psi$-loops.

\subsection{Constructing X-cube model via $e$-loop condensation}
\label{sec: xcube from dZ2}
Here, we reconstruct the XC model (on truncated cubic lattice) from the decoupled layers of $\mathbb{Z}_2$ HGW model (d$\mathbb{Z}_2$ for short) by condensing the $e$-loops in our framework. Each principal edge carries a label taking value in the set:
\begin{equation}
    L_{\mathrm{d}\mathbb{Z}_2}=\{j_E | k_E \mid j_E,k_E\in L_{\mathbb{Z}_2}=\{1, \psi\}\}=\{1|1,1|\psi,\psi|1,\psi|\psi\}.
    \label{DZ2 dof}
\end{equation}
Only one subset of $L_{\mathrm{d} \mathbb{Z}_2}$ is closed under $\mathbb{Z}_2$ fusion rules:
\begin{equation}
    L_\mathcal{S}=\{1|1,\psi|\psi\}:=L_\text{XC},
    \label{XC dof}
\end{equation}
in which the degree of freedom on a principal edge of the XC model takes value. There is only one chargeon creation operators in the d$\mathbb{Z}_2$ model: $W^{e;1,1}$. Note that to project \eqref{DZ2 dof} to \eqref{XC dof}, we need to gap out gauge degree of freedom values $1\psi$ and $\psi 1$. The unique p-loop condensation projector that does the job is $P^{\text{XC} \mid \mathrm{d} \mathbb{Z}_2}$: 
\begin{equation}
    P^{\text{XC} \mid \mathrm{d} \mathbb{Z}_2}=\prod_E P^{\text{XC} \mid \mathrm{d} \mathbb{Z}_2}_E, \quad P^{\text{XC} \mid \mathrm{d} \mathbb{Z}_2}_E=\frac{W^{1;1,1}_E V^{1;1,1}_E + W^{e;1,1}_E V^{e;1,1}_E }{2}.
    \label{eq: e loop projector}
\end{equation}

\paragraph{Effective Hamiltonian} 

The effective Hamiltonian $H^\text{eff}_\text{child}$ of the child model can be obtained from the partition function  $Z_{\text{child}}$ of the child model, which can obtained by projecting the partition function of the parent model $Z_{\text{parent}}$:
\begin{equation}
    Z_{\text{child}} = \mathrm{Tr}(P Z_{\text{parent}} P), \quad Z_{\text{parent}}=\exp(-\beta H_{\text{parent}}),
\end{equation}
where $P$ is the global projector $P=\prod_E P_E$ acting on all principal edges, and $H_{\text{parent}}$ is the parent Hamiltonian. Then effective child Hamiltonian $H^\text{eff}_\text{child}$ can be read from $Z_{\text{child}}$.

In the X-cube case, the global projector is $P^{\text{XC} \mid \mathrm{d} \mathbb{Z}_2}$ defined in \eqref{eq: e loop projector}, and the parent Hamiltonian is 
\begin{equation}
    H_{\text{parent}}=H_{\mathrm{d} \mathbb{Z}_2}=\sum_{(E,\mu)} H_{\mathbb{Z}_2}^{(E,\mu)}.
\end{equation}
After projection and series expansion, we find the effective child Hamiltonian is
\begin{equation}
   H^\text{eff}_\text{child} = -\sum_{(E,\mu)}\left(\sum_{P_d} Q_{P_d} + \sum_{P_o} Q_{P_o}^1\right) - K_C \sum_C F_C:=H_\text{XC}, \quad F_C=\prod_{P_o \in \partial C}Q_{P_o}^\psi,
\end{equation}
where $F_C$ is the cube operator of the XC model and $K_C$ is the constant factor that can be computed if desired. This Hamiltonian is exactly the X-cube Hamiltonian defined on the truncated cubic lattice. We also note that this result matches the effective Hamiltonian obtained via Brillouin-Wigner perturbation theory in \cite{ma_fracton_2017}.

\paragraph{Quasiparticle spectrum}
There are four anyon types in the $\mathbb{Z}_2$ HGW string-net model: the trivial anyon $1$ with a unique dyonic sector $(1,1)$, the pure fluxon $m$ with a unique dyonic sector $(m,\psi)$, the pure chargeon $e$ with a unique dyonic sector $(e,1)$, and  $\epsilon=(e \times m)$ with a unique dyonic sector $(\epsilon,\psi)$.

Consider projecting the trivial anyon of the $\mathbb{Z}_2$ HGW string-net model. Acting a local projector $P^{\text{XC} \mid \mathrm{d} \mathbb{Z}_2}_E$ (defined in \eqref{eq: e loop projector}) on a principal edge $E_1$ returns\footnote{For simplicity, we do not draw secondary edges in the truncated cubic lattices in \eqref{eq: p-loop act 1} \eqref{eq: p-loop act 2}.}: 

\begin{equation}
    P^{\text{XC} \mid \mathrm{d} \mathbb{Z}_2}_{E_1} \vcenter{\hbox{\includegraphics[height=12ex]{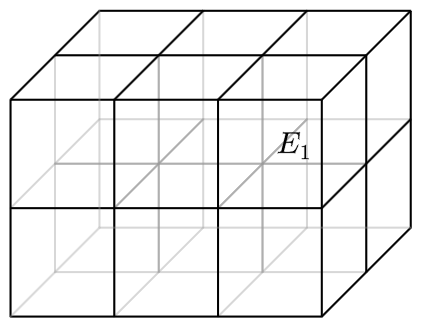}}} = \vcenter{\hbox{\includegraphics[height=12ex]{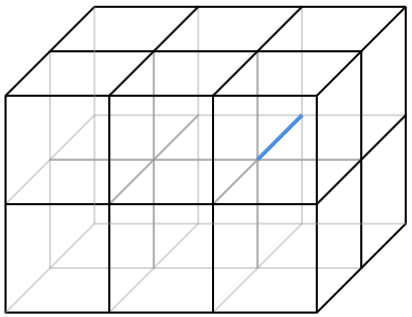}}} + \vcenter{\hbox{\includegraphics[height=12ex]{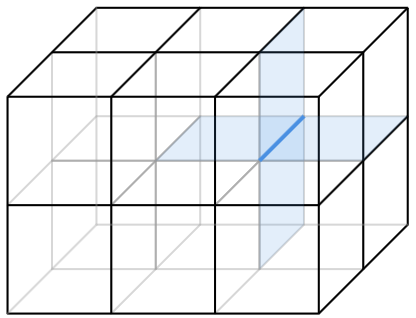}}},
    \label{eq: p-loop act 1}
\end{equation}
where the thick blue line in each term highlights the principal edge $E_1$, and a light blue square stands for an $e$ excitation (4 in total). The four $e$ excitations form a smallest $e$-loop. Subsequently, acting $P^{\text{XC} \mid \mathrm{d} \mathbb{Z}_2}_E$ on a neighboring principal edge $E_2$ will result in:
\begin{equation}
\begin{split}
    P^{\text{XC} \mid \mathrm{d} \mathbb{Z}_2}_{E_2} P^{\text{XC} \mid \mathrm{d} \mathbb{Z}_2}_{E_1} \vcenter{\hbox{\includegraphics[height=12ex]{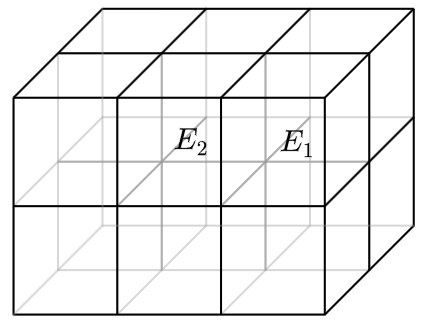}}} = &\vcenter{\hbox{\includegraphics[height=12ex]{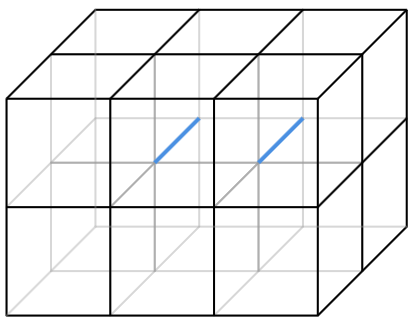}}} + \vcenter{\hbox{\includegraphics[height=12ex]{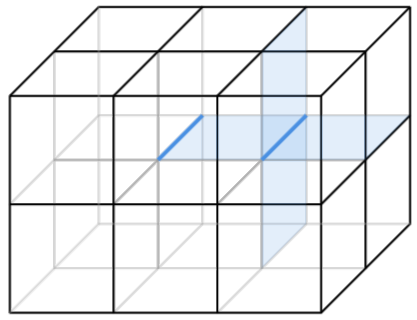}}} + \\ &\vcenter{\hbox{\includegraphics[height=12ex]{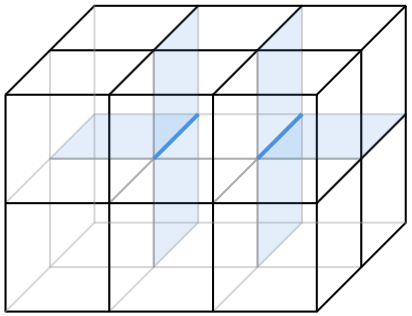}}} + \vcenter{\hbox{\includegraphics[height=12ex]{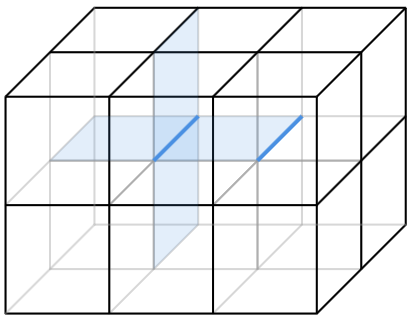}}},
    \label{eq: p-loop act 2}
\end{split}    
\end{equation}
where larger $e$-loops appear. We can continue this procedure to apply the local projects to all principal edges and obtain the action of the total projector $P^{\text{XC} \mid \mathrm{d} \mathbb{Z}_2}$ on the whole lattice. The resultant state is a coherent state of $e$-loops. For simplicity, we use the following shorthand expression to express the state:
\begin{equation}
    P^{\text{XC} \mid \mathrm{d} \mathbb{Z}_2}  \vcenter{\hbox{\includegraphics[height=12ex]{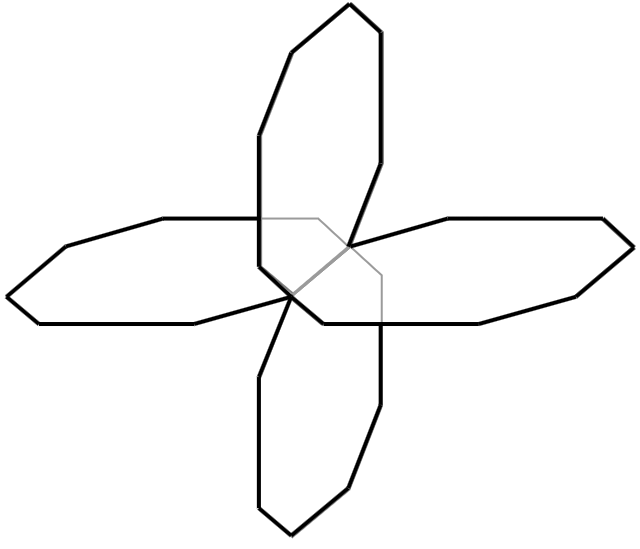}}} = \vcenter{\hbox{\includegraphics[height=12ex]{dl_vacuum_simp.png}}} + \vcenter{\hbox{\includegraphics[height=12ex]{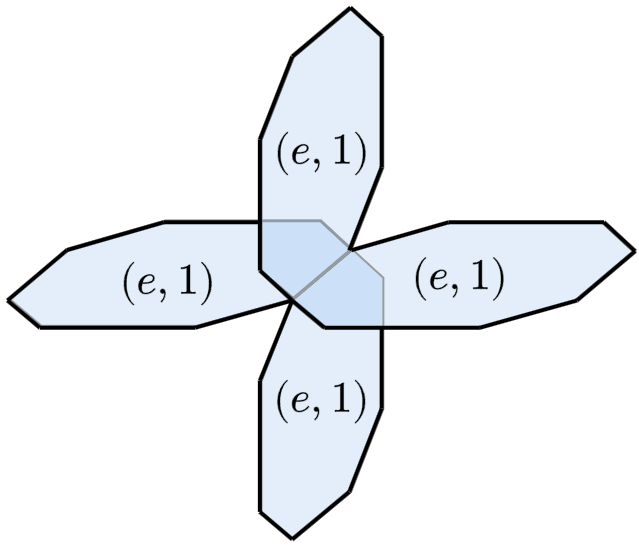}}} + \ldots,
    \label{eq: xcube vac}
\end{equation}
where only a star-shape part of the whole lattice is shown, and the ellipse includes all the other possible terms in the coherent state.



In the XC model, there are one fracton type $\mathfrak{f}_e$ and two lineon types $\mathfrak{l}_{mm}$ and $\mathfrak{l}_{m \epsilon}$. If there is an odd number of $(e,1)$ on the octagonal faces of a cube $C$, we say that the cube $C$ supports a fracton excitation $\mathfrak{f}_e$\cite{ma_fracton_2017}. The configuration of a pair of $(e,1)$ on the two sides of a principal edge $E$ in the same plane in the d$\mathbb{Z}_2$ model is always projected into a state containing four fractons in the XC model. For example, consider an isolated pair\footnote{Other excitations must be at least two plaquettes away in all directions.} of $(e,1)$ in the $\mathrm{d}\mathbb{Z}_2$ model, we have 
\begin{equation}
    P^{\text{XC} \mid \mathrm{d} \mathbb{Z}_2} \vcenter{\hbox{\includegraphics[height=12ex]{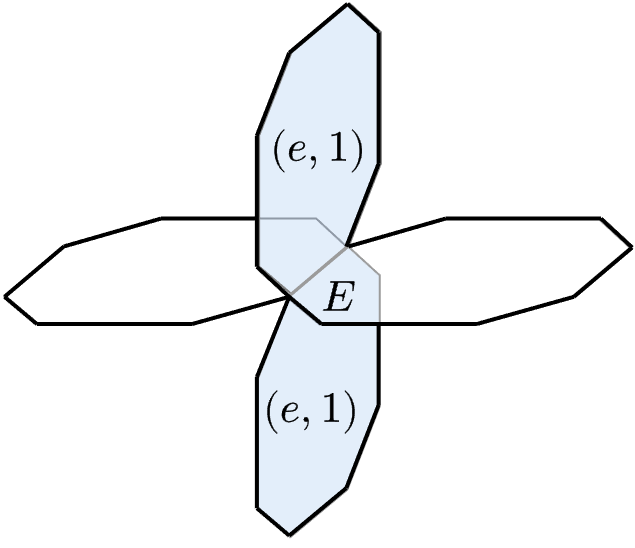}}} 
    =P^{\text{XC} \mid \mathrm{d} \mathbb{Z}_2} \vcenter{\hbox{\includegraphics[height=12ex]{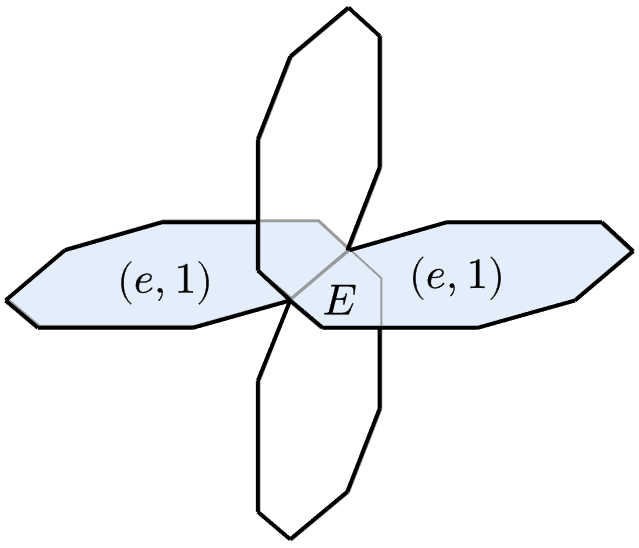}}} 
    =\vcenter{\hbox{\includegraphics[height=12ex]{fracton_e_1.png}}} +\vcenter{\hbox{\includegraphics[height=12ex]{fracton_e_2.png}}} + \ldots,
\end{equation}
where the right hand side is the state containing four fractons $\mathfrak{f}_{e}$ in the four cubes around the principal edge $E$ in the XC model\cite{ma_fracton_2017}.

Planar configurations of any pair of dyons with a $\psi$ flux, such as $(m,\psi)$-pair and $(\epsilon,\psi)$-pair, are projected out of the XC Hilbert space. For example,
\begin{equation}
    P^{\text{XC} \mid \mathrm{d}\mathbb{Z}_2} \vcenter{\hbox{\includegraphics[height=12ex]{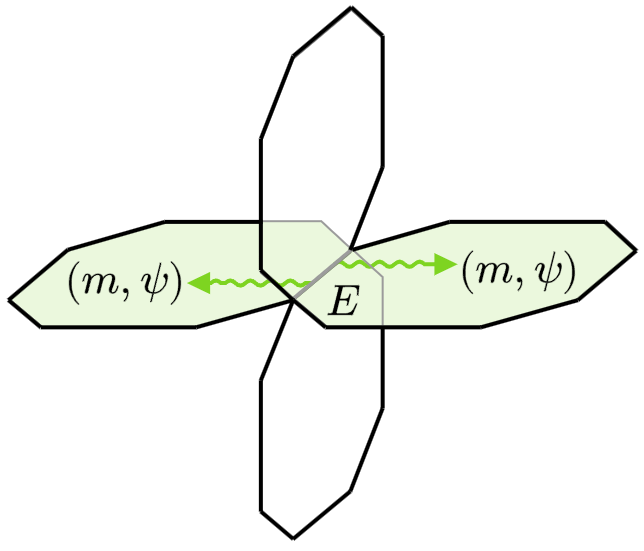}}}=P^{\text{XC} \mid \mathrm{d}\mathbb{Z}_2} \vcenter{\hbox{\includegraphics[height=12ex]{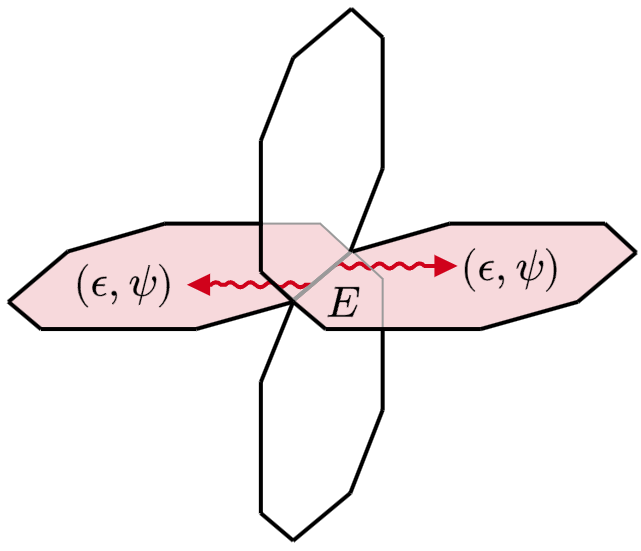}}}=0. \label{eq:confinement_zero}
\end{equation}
The above equation indicates that anyons with $\psi$ fluxes in an isolated planar configuration are confined. This confinement is physically attributed to the non-trivial braiding between these anyons with $\psi$ fluxes and the condensed $e$-loop in the child phase. Nevertheless, when these anyons form bound states in a ``star configuration'' (formed by two perpendicular pairs of dyons, each carrying a $\psi$ flux. See the left hand side of \eqref{lineon m m} for example), the combination of two anyons with $\psi$ flux have trivial braiding with $e$-loops, rendering these combined excitations to be de-confined in the XC Hilbert space. These bound states are lineons\cite{ma_fracton_2017}.

\begin{figure}[ht]
  \centering
  \includegraphics[width=\linewidth]{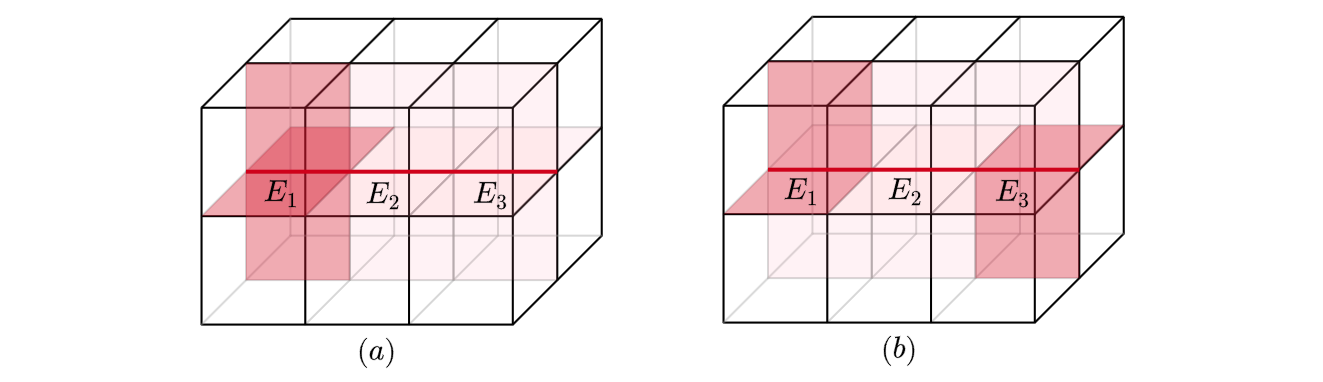}
  \caption{Restricted one-dimensional mobility of lineon. The thick red line denotes the strictly 1D linear subsystem along which the lineons are confined to move. Red plaquettes represent anyons carrying a $\psi$-flux, and the light red planes indicate the respective planes within which these $\psi$-fluxes reside. (a) A configuration where two lineons are localized on the principal edge $E_1$. (b) The configuration after one of the lineons has hopped from edge $E_1$ along the restricted 1D line to edge $E_3$.}
  \label{fig:xcube_lineon}
\end{figure}

A lineon is an excitation whose mobility is strictly restricted to that 1D linear subsystem\cite{vijay_fracton_2016,vijay2017isotropiclayerconstructionphase}. In our extended cage-net construction, this one-dimensional restricted mobility is realized on a straight line consisting of principal edges when it supports two perpendicular $\psi$ tails (see Fig. \ref{fig:xcube_lineon} as an example); we refer to this specific configuration as a lineon excitation $\mathfrak{l}$.

Upon applying the global projector, an isolated star configuration around a principal edge is projected into a state consisting of two lineons in the XC model. For example, the star configuration of two perpendicular $(m,\psi)$-pairs around a principal edge $E$ is projected into a state consisting of two lineons, denoted as $\mathfrak{l}_{mm}$ (where subscript $mm$ denotes the anyon types in the horizontal and vertical planes):
\begin{equation}
    P^{\text{XC} \mid \mathrm{d} \mathbb{Z}_2}\vcenter{\hbox{\includegraphics[height=12ex]{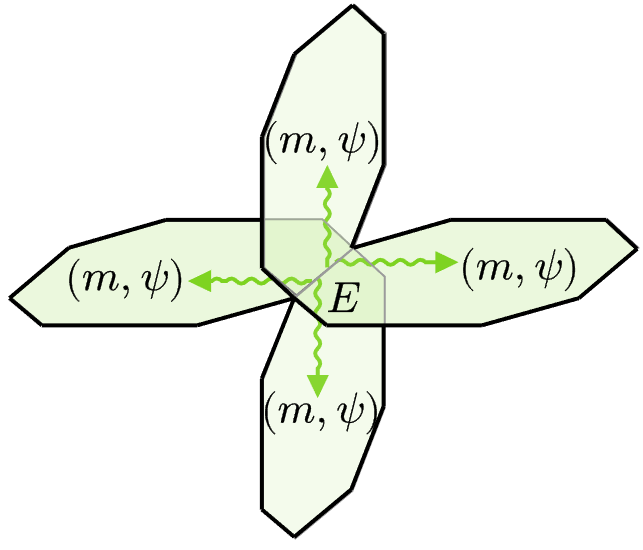}}} =\vcenter{\hbox{\includegraphics[height=12ex]{lineon_mm_1.png}}} +\vcenter{\hbox{\includegraphics[height=12ex]{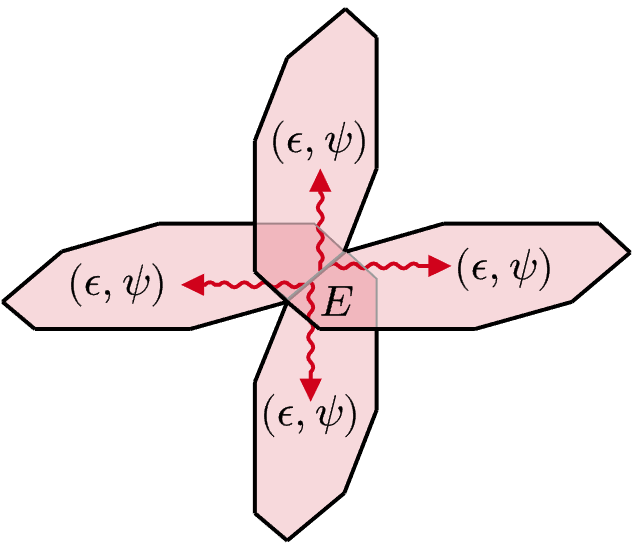}}} + \ldots.
    \label{lineon m m}
\end{equation}
Because the new vacuum of the XC model is a condensate of $e$-loops, excited states in the parent model that differ only by the fusion of an $e$-loop become physically indistinguishable in the child phase. For example, the star configuration of two perpendicular $(\epsilon,\psi)$-pairs -- which can be converted into the aforementioned $(m,\psi)$ star configuration by fusing a $e$-loop -- is projected into the same state:

\begin{equation}
    P^{\text{XC} \mid \mathrm{d} \mathbb{Z}_2}\vcenter{\hbox{\includegraphics[height=12ex]{lineon_mm_2.png}}} =\vcenter{\hbox{\includegraphics[height=12ex]{lineon_mm_1.png}}} +\vcenter{\hbox{\includegraphics[height=12ex]{lineon_mm_2.png}}} + \ldots.
    \label{lineon m m}
\end{equation}

Similarly, we have the following projection:
\begin{equation}
    P^{\text{XC} \mid \mathrm{d} \mathbb{Z}_2}\vcenter{\hbox{\includegraphics[height=12ex]{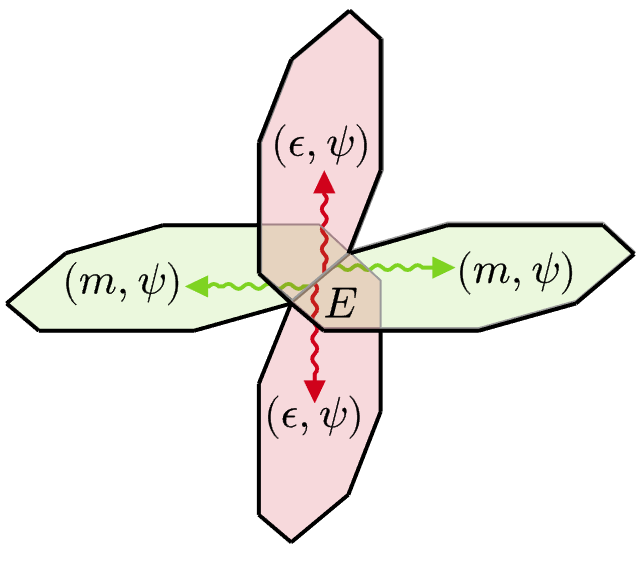}}} 
    =P^{\text{XC} \mid \mathrm{d} \mathbb{Z}_2}\vcenter{\hbox{\includegraphics[height=12ex]{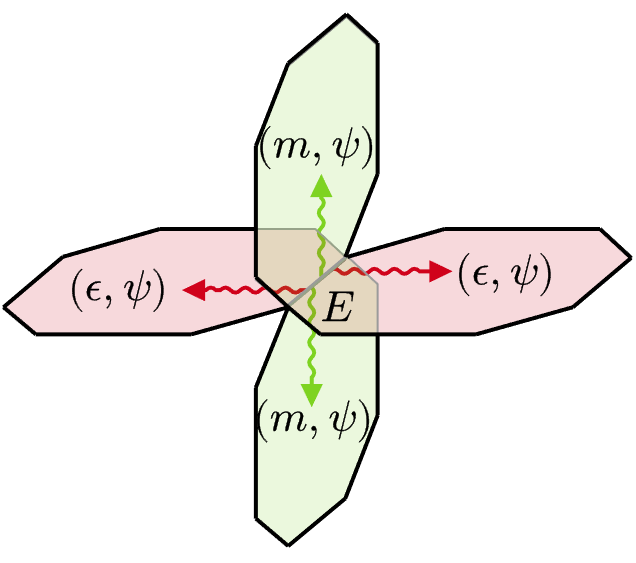}}}
    =\vcenter{\hbox{\includegraphics[height=12ex]{lineon_m_ep_1.png}}} +\vcenter{\hbox{\includegraphics[height=12ex]{lineon_m_ep_2.png}}} + \ldots,
\end{equation}
where the right hand side is precisely the state consisting two lineons $\mathfrak{l}_{m \epsilon}$.

\subsection{Constructing extended ICN model via $\psi\Bar{\psi}$-loop condensation}

Now we consider condensing $\psi\Bar{\psi}$-loops in the decoupled layers of Ising HGW model (dDI for short). The largest subset $L_\mathcal{S}$ of the set 
\begin{equation}
    L_\text{dDI}=\{j_E | k_E \in L_\text{DI} \mid L_\text{DI}=1,\psi,\sigma \}=\{1|1,\psi |\psi ,\sigma| \sigma ,1|\psi ,\psi| 1\, 1|\sigma, \sigma|1, \sigma|\psi,\psi|\sigma\}
\end{equation}
that is closed under the fusion rules of Ising UFC is:
\begin{equation}
    L_{\text{ICN}}:=L_\mathcal{S}=\{1|1,\psi| \psi ,\sigma| \sigma ,1|\psi ,\psi| 1\},
\end{equation}
which is the label set for the principal-edge degree of freedom of the ICN model. 

Using the eigenvalues of the chargeon creation operators (detailed in Appendix \ref{doubled ising data}), the unique p-loop condensation operator that gaps out all the elements not in $L_{\text{ICN}}$ is
\begin{equation}
    P^{\text{ICN} \mid \text{dDI}}=\prod_E {P^{\text{ICN} \mid \text{dDI}}_E} , \quad P^{\text{ICN} \mid \text{dDI}}_E=\frac{W^{1\Bar{1} ;1,1}_E V^{1\Bar{1} ;1,1}_E +W^{\psi \Bar{\psi } ;1,1}_E V^{\psi \Bar{\psi } ;1,1}_E}{2},
    \label{psibarpsi-loop projector}
\end{equation}
whose matrix form matches that of the coupling term in \cite{prem_cage-net_2019}, which may realize $\psi\Bar{\psi}$-loop condensation. 

\paragraph{Effective Hamiltonian} The effective child Hamiltonian obtained by $\psi\bar{\psi}$-loop condensation is:
\begin{equation}
    H_\text{ICN}:=H_\text{child}^\mathrm{eff} =-\sum_{(E,\mu)}\sum_{K_d}{Q_{P_d}}-K_{P}\sum_{(E,\mu)}\sum_{P_o} \left( {Q^1_{P_o}}+{Q^{\psi}_{P_o}}\right)-K_C\sum_C{F_C},
\end{equation}
where $K_P$ and $K_C$ are the constant factors that can be computed if desired. The operator $F_C$ in the equation above is the cube operator of the ICN model:
\begin{equation}
    F_C=\prod_{P_o \in C} Q^{\sigma}_{P_o},
\end{equation}
where $Q^{\sigma}_{P_o}$ is defined in Appendix \ref{HGW theory}. This effective Hamiltonian agrees with the ICN Hamiltonian obtained in \cite{prem_cage-net_2019}.

\subsection{Quasiparticle types in ICN model}
There are eleven quasiparticle types in the ICN model:
\begin{equation}
\begin{array}{llll}
    \text{Fracton:}\quad &\mathfrak{f}_{\psi\bar{\psi}}, & &\\[4pt]
    \text{Planon:\ } &\mathfrak{p}_{1\bar{\psi}}, &\mathfrak{p}_{\psi\bar{1}}, &\mathfrak{p}_{\sigma\bar{\sigma}},\\[3pt]
    \text{Lineon:\ } &\mathfrak{l}_{\sigma\bar{1},\sigma\bar{1}}, \ \ \ &\mathfrak{l}_{\sigma\bar{1},\sigma\bar{\psi}},\ \ \  &\mathfrak{l}_{1 \bar\sigma ,1 \bar\sigma},\quad \mathfrak{l}_{1 \bar\sigma ,\psi \bar\sigma},\quad 
    \mathfrak{l}_{1 \bar\sigma ,\sigma\bar{1}},\quad \mathfrak{l}_{1\bar\sigma,\sigma\bar\psi}.
\end{array}
\label{ICN quasiparticle types}
\end{equation}
These quasiparticle types can be obtained respectively by projecting appropriate configurations of anyons in the dDI model via projector $P^{\text{ICN} \mid \text{dDI}}$ \eqref{psibarpsi-loop projector}. In what follows, we shall showcase a few examples.

The configuration of an isolated pair of $(\psi\bar\psi,1)$ on the two sides of a principal edge in the same plane in the dDI model is always projected into a state containing four fractons in the ICN model. For example, 

\begin{equation}
    P^{\text{ICN} \mid \text{dDI}} \vcenter{\hbox{\includegraphics[height=12ex]{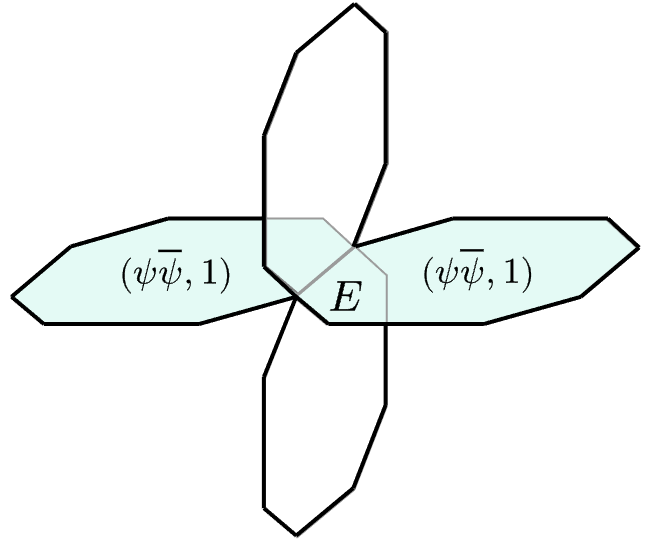}}} 
    =P^{\text{ICN} \mid \text{dDI}} \vcenter{\hbox{\includegraphics[height=12ex]{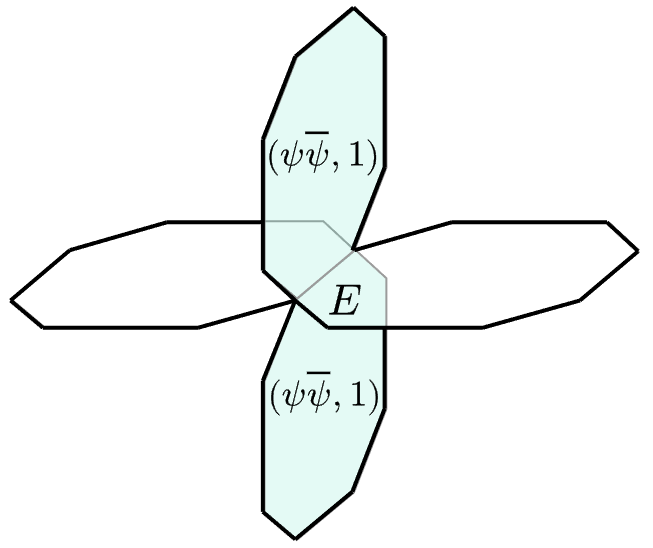}}} 
    =\vcenter{\hbox{\includegraphics[height=12ex]{psibarpsi-fracton_left.png}}} +\vcenter{\hbox{\includegraphics[height=12ex]{psibarpsi-fracton_right.png}}} + \ldots,
    \label{eq: ICN fracton}
\end{equation}
where the right hand side meets the definition of a fracton state in the ICN model\cite{prem_cage-net_2019}.

An isolated planar configuration of  $(1\bar{\psi},\psi)$ in the dDI model is projected into a state consisting two planons $\mathfrak{p}_{1\bar{\psi}}$:
\begin{equation}
    P^{\text{ICN} \mid \text{dDI}}\vcenter{\hbox{\includegraphics[height=12ex]{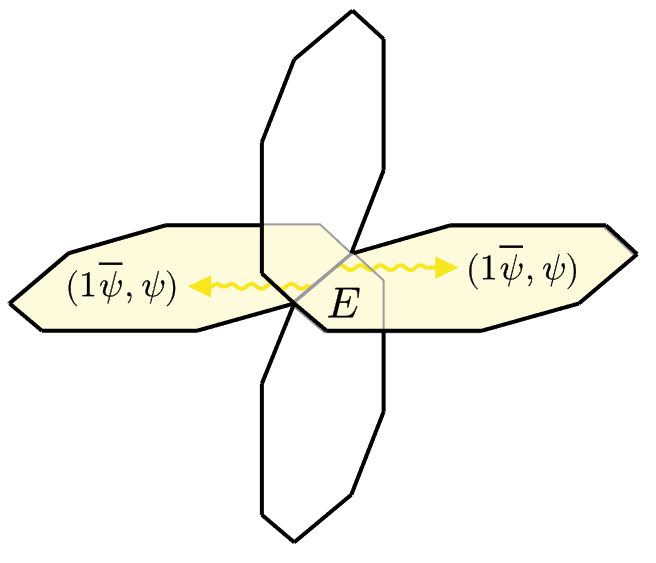}}} 
    =\vcenter{\hbox{\includegraphics[height=12ex]{1barpsi_vac.png}}} +\vcenter{\hbox{\includegraphics[height=12ex]{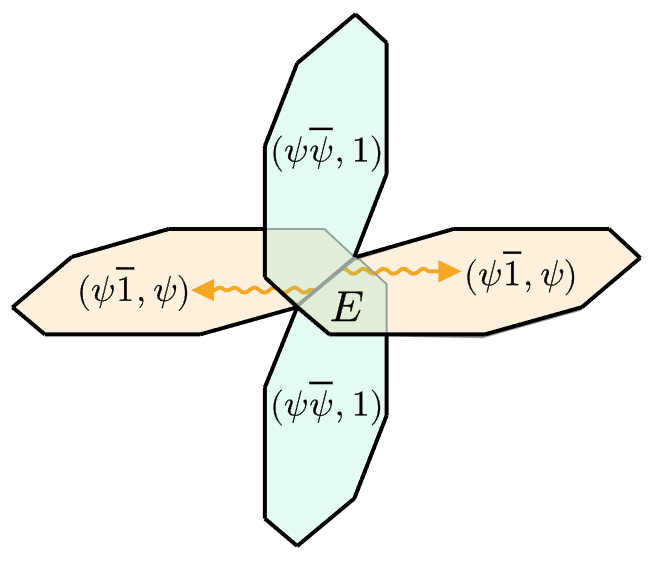}}} + \ldots,
\end{equation}
where the two planons reside in the two horizontal plaquettes. These quasiparticles are called planons because they are free to move within the horizontal plane.

Similarly, an isolated planar configuration of  $(\psi\bar{1},\psi)$ in the dDI model is projected into a state consisting two planons $\mathfrak{p}_{\psi\bar{1}}$:

\begin{equation}
    P^{\text{ICN} \mid \text{dDI}}\vcenter{\hbox{\includegraphics[height=12ex]{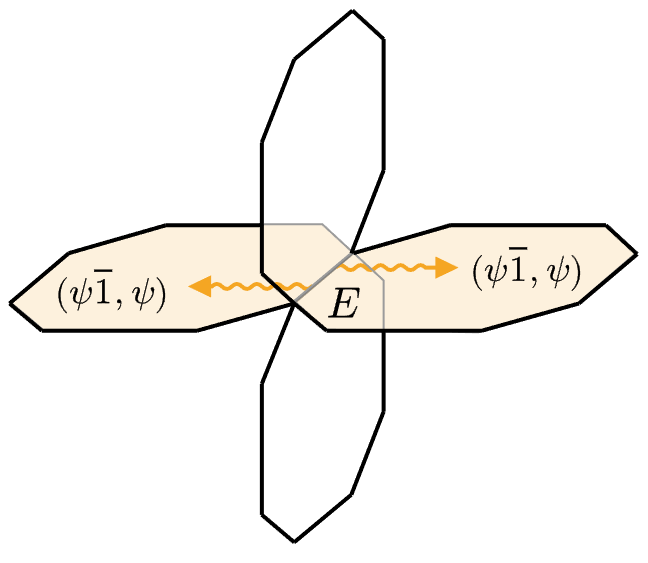}}} 
    =\vcenter{\hbox{\includegraphics[height=12ex]{psibar1_vac.png}}} +\vcenter{\hbox{\includegraphics[height=12ex]{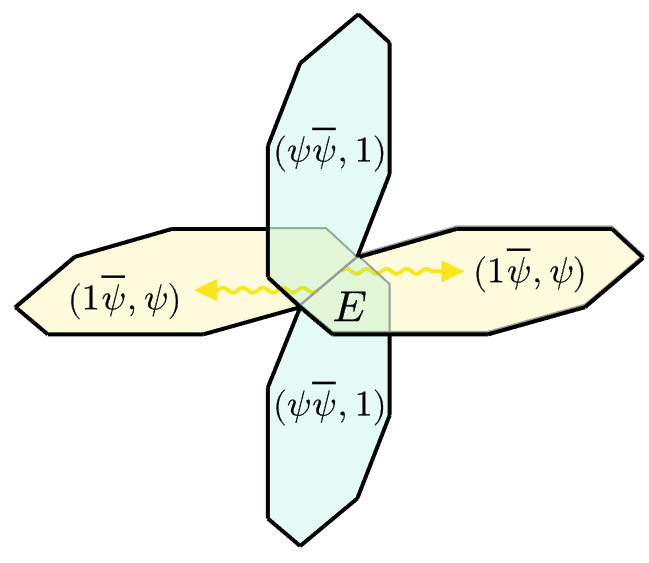}}} + \ldots.
\end{equation}

Likewise, isolated planar configurations of a pair of $\sigma\bar{\sigma}$ are projected into states consisting two planons $\mathfrak{p}_{\sigma\bar{\sigma}}$. There are two internal sectors (unobservable) within planon $\mathfrak{p}_{\sigma\bar{\sigma}}$, which are $\mathfrak{p}_{(\sigma\bar{\sigma},1)}$ and $\mathfrak{p}_{(\sigma\bar{\sigma},\psi)}$:
\begin{equation}
    P^{\text{ICN} \mid \text{dDI}}\vcenter{\hbox{\includegraphics[height=12ex]{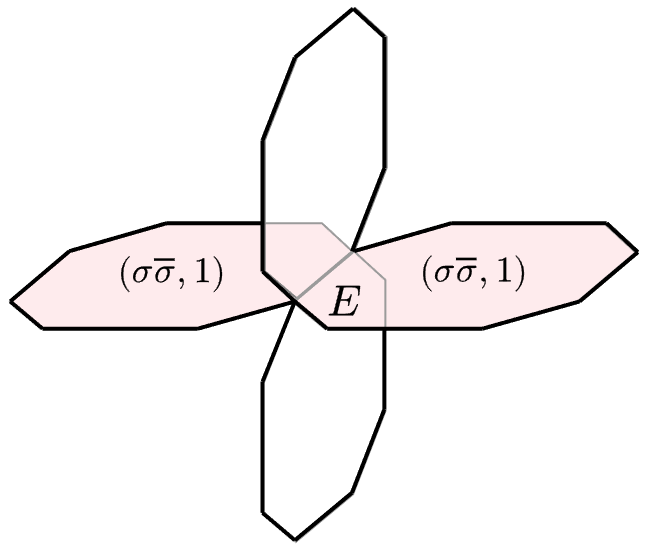}}} 
    =\vcenter{\hbox{\includegraphics[height=12ex]{sigbarsig_vac.png}}} +\vcenter{\hbox{\includegraphics[height=12ex]{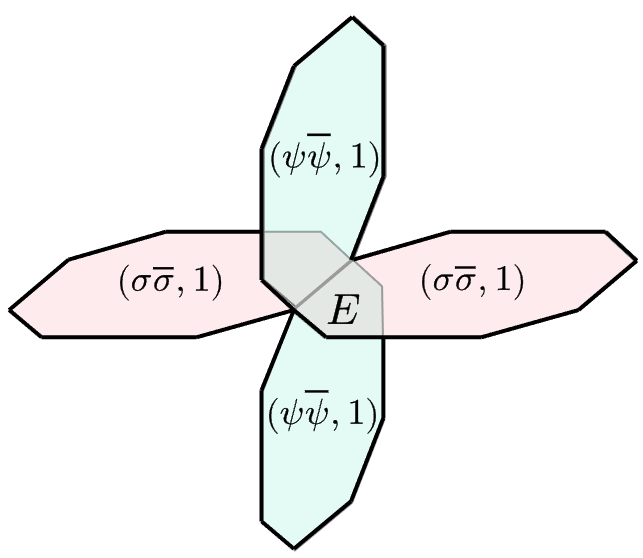}}} + \ldots,
\end{equation}

\begin{equation}
    P^{\text{ICN} \mid \text{dDI}}\vcenter{\hbox{\includegraphics[height=12ex]{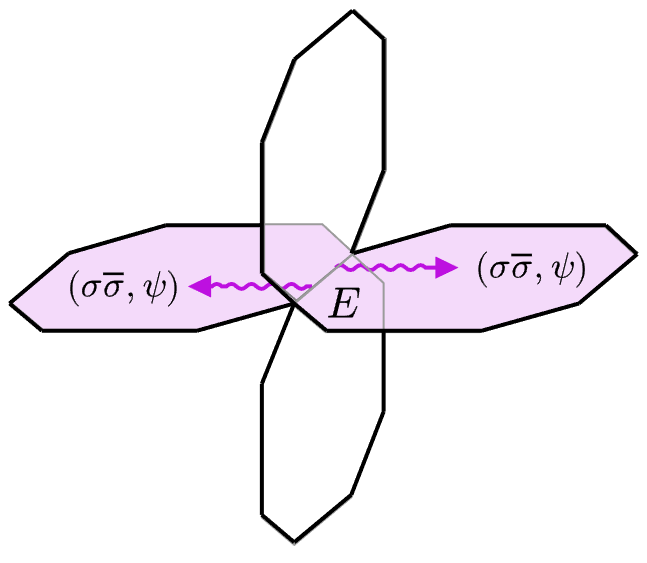}}} 
    =\vcenter{\hbox{\includegraphics[height=12ex]{sigbarsig,psi_vac.png}}} +\vcenter{\hbox{\includegraphics[height=12ex]{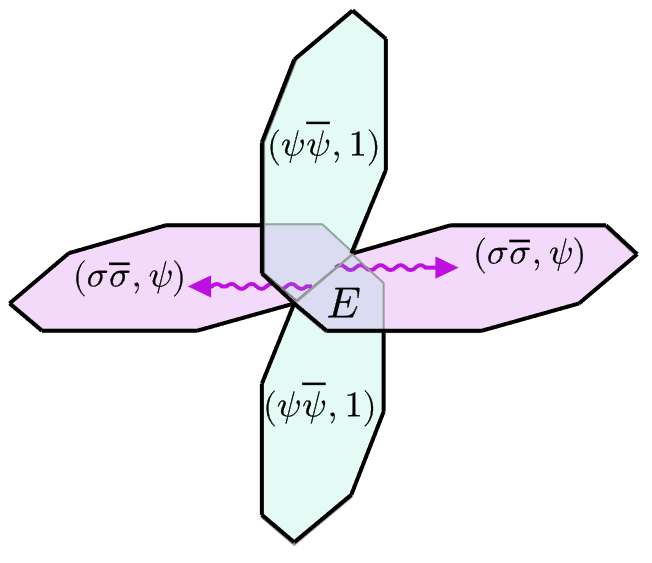}}} + \ldots,
\end{equation}
where the right hand sides are the states corresponding to the two planon sectors of $\mathfrak{p}_{\sigma\bar{\sigma}}$ residing in the two horizontal plaquettes.

Isolated planar configurations of any pair of anyons with a $\sigma$ flux, such as $(\sigma\bar{1},\sigma)$-pair and $(\sigma\bar{\psi},\sigma)$-pair, are projected out of the ICN Hilbert space. For example,
\begin{equation}
   P^{\text{ICN} \mid \text{dDI}} \vcenter{\hbox{\includegraphics[height=12ex]{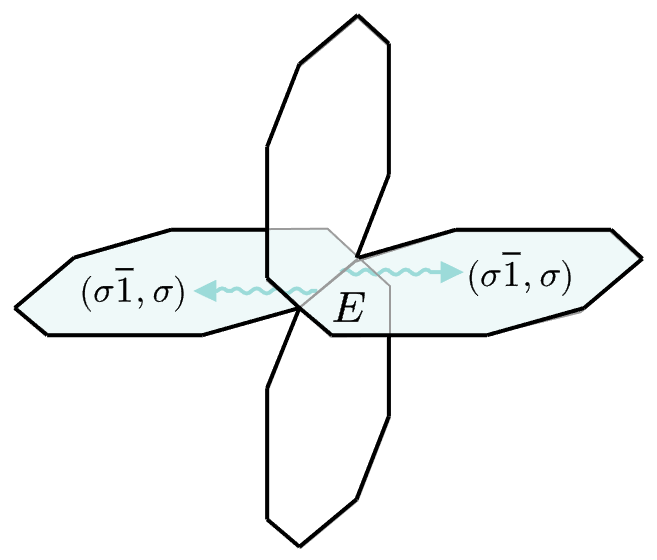}}}=P^{\text{ICN} \mid \text{dDI}} \vcenter{\hbox{\includegraphics[height=12ex]{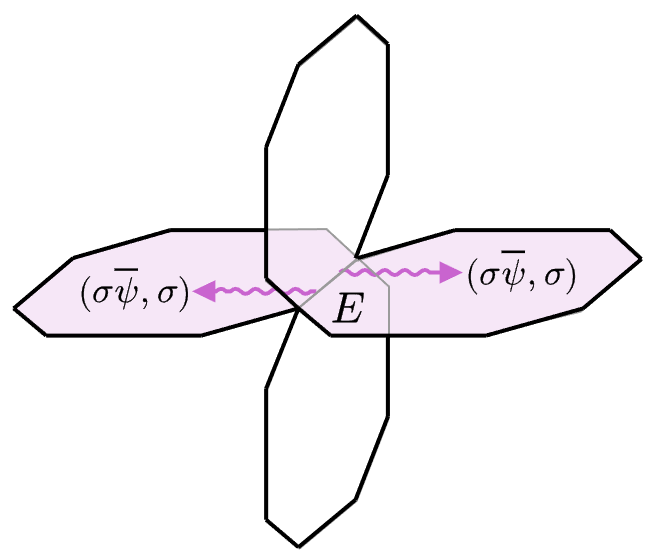}}}=0.
\end{equation}

Nonetheless, an isolated star configuration of two perpendicular pairs of anyons with a $\sigma$ flux around a principal edge is projected to a state consisting two lineons along an edge in the ICN model. If two such star configurations only differ by fusing a $\psi\bar{\psi}$-loop, they are projected into the same state. For instance, the isolated star configuration of two perpendicular pairs of $(\sigma\bar{1},\sigma)$ and that of two perpendicular pairs of $(\sigma\bar{\psi},\sigma)$ are projected into the same state consisting two lineons $\mathfrak{l}_{\sigma\bar{1},\sigma\bar{1}}$ in the ICN model:
\begin{equation}
    P^{\text{ICN} \mid \text{dDI}}\vcenter{\hbox{\includegraphics[height=12ex]{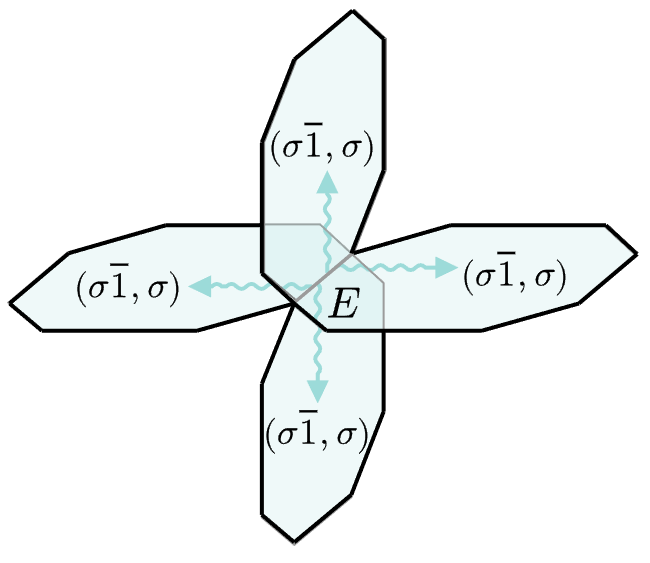}}} 
    =P^{\text{ICN} \mid \text{dDI}}\vcenter{\hbox{\includegraphics[height=12ex]{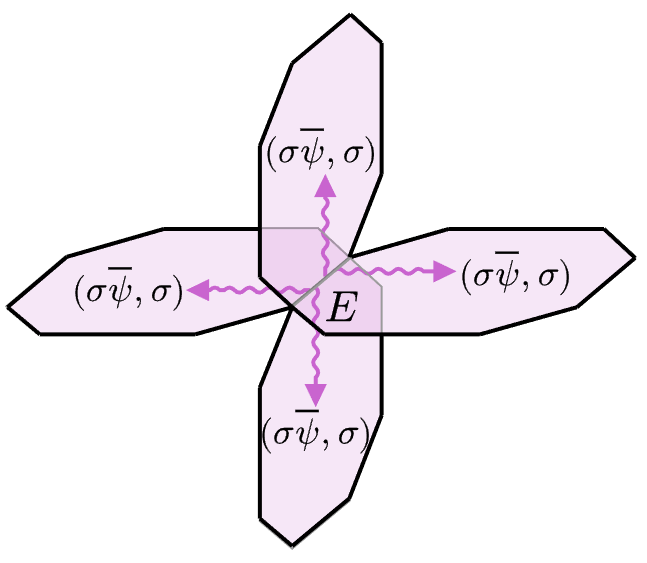}}}
    =\vcenter{\hbox{\includegraphics[height=12ex]{sigbar1-loop.png}}} +\vcenter{\hbox{\includegraphics[height=12ex]{sigbarpsi-loop.png}}} + \ldots.
    \label{lineon sig sig}
\end{equation}

Similarly, we have the following projection:
\begin{equation}
    P^{\text{ICN} \mid \text{dDI}}\vcenter{\hbox{\includegraphics[height=12ex]{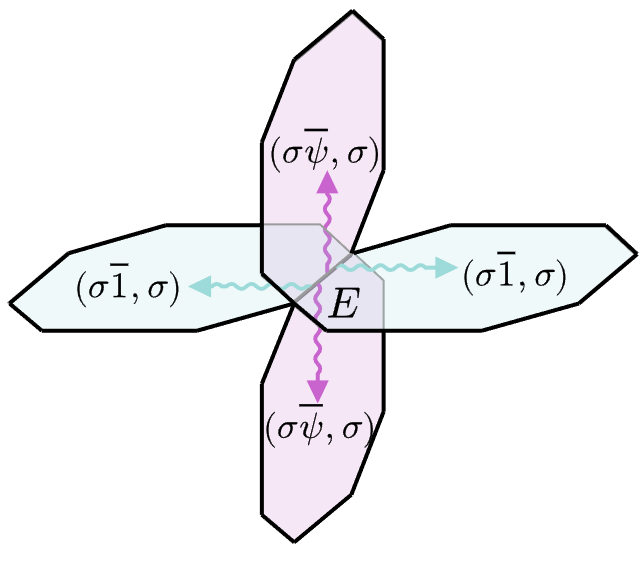}}} 
    =P^{\text{ICN} \mid \text{dDI}}\vcenter{\hbox{\includegraphics[height=12ex]{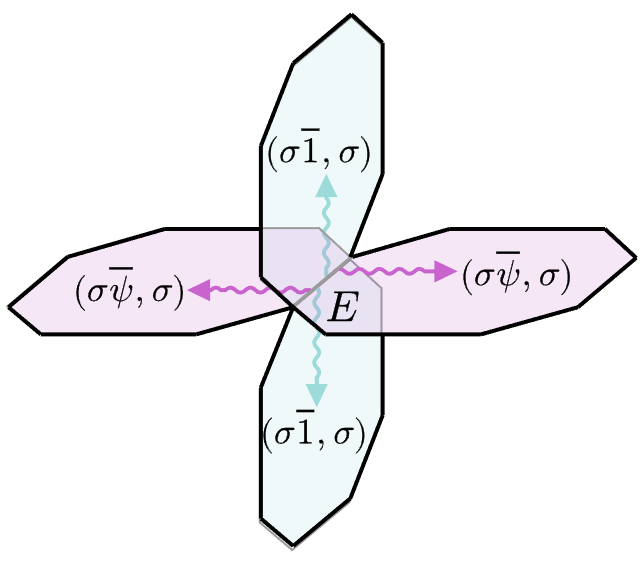}}}
    =\vcenter{\hbox{\includegraphics[height=12ex]{sigbar1_sigbarpsi.png}}} +\vcenter{\hbox{\includegraphics[height=12ex]{sigbarpsi_sigbar1.png}}} + \ldots,
\end{equation}
where the right hand side is the state consisting two lineons $\mathfrak{l}_{\sigma\bar{1},\sigma\bar{\psi}}$.

\section{Fracton condensation and decoupling}
\label{fracton cond}

In this section, we discuss fracton condensation in extended CN models. To be specific, we take the extended ICN model as an example and condense fracton $\mathfrak{f}_{\psi\bar{\psi}}$ therein\footnote{Fracton condensation in the XC model has been discussed in \cite{PhysRevB.104.165121, zhu2023}.}. Based on the results in \eqref{eq: ICN fracton}, we are now able to write down the creation operator of $\mathfrak{f}_{\psi\bar{\psi}}$ fractons $C^{1\bar{1},\psi\bar{\psi}}_E$,
\begin{equation}
   C^{1\bar{1},\psi\bar{\psi}}_E=\frac{1}{2}\left( W^{1\Bar{1} ;1,1}_E V^{\psi \Bar{\psi } ;1,1}_E +W^{\psi \Bar{\psi } ;1,1}_E V^{1\Bar{1} ;1,1}_E \right). 
   \label{eq: fracton creation}
\end{equation}
Using these creation operators, we can define the condensation operator of fracton $\mathfrak{f}_{\psi\bar{\psi}}$ as
\begin{equation}
    P^{\mathrm{d}\mathbb{Z}_2 \mid \text{ICN}}=\prod_E \left( P^{\mathrm{d}\mathbb{Z}_2 \mid \text{ICN}}_E \right), \quad P^{\mathrm{d}\mathbb{Z}_2 \mid \text{ICN}}_E = \frac{1}{2}\left(\mathbf{1}+C^{1\bar{1},\psi\bar{\psi}}_E \right).
\end{equation}
The projector $P^{\mathrm{d}\mathbb{Z}_2 \mid \text{ICN}}$ gaps out the degree of freedom $\sigma|\sigma$ from $L_\text{ICN}$, and we obtain the label set of $\mathrm{d}\mathbb{Z}_2$ after the fracton condensation:
\begin{equation}
    L_{\mathrm{d}\mathbb{Z}_2} = \{1|1, 1|\psi, \psi| 1, \psi| \psi\} = \{j_E| k_E \mid j_E,k_E \in L_{\mathbb{Z}_2}\}.
    \label{eq: LdZ2}
\end{equation}

Since the lineons in the extended ICN model carry a $\sigma$-flux, they are projected out under the fracton condensation. The two internal sectors $\mathfrak{p}_{(\sigma\bar{\sigma},1)}$ and $\mathfrak{p}_{(\sigma\bar{\sigma},\psi)}$ of planon $\mathfrak{p}_{\sigma\bar{\sigma}}$ split and are projected differently. The $\mathfrak{p}_{(\sigma\bar{\sigma},1)}$ sector is projected to the $e$ anyon in the $\mathbb{Z}_2$ toric code layers, while the $\mathfrak{p}_{(\sigma\bar{\sigma},\psi)}$ sector is projected to the $m$ anyon. The different behavior of the two internal sectors of planon $\mathfrak{p}_{\sigma\bar{\sigma}}$ in the ICN model upon $(\sigma\bar{\sigma},1)$-loop condensation is a 3-dimensional counterpart of the splitting of anyons occuring in the scenario of anyon condensation in two-dimensional topological orders\cite{bais_condensate-induced_2009,hung_generalized_2015,hu_anyon_2022}. We manifest such a splitting at the level of states in our extended ICN model, as is done for anyon condensation in the HGW string-net model\cite{wang_electric-magnetic_2020,hu_electric-magnetic_2020,zhao_characteristic_2023}.

Furthermore, the planons $\mathfrak{p}_{1\bar{\psi}}$ and $\mathfrak{p}_{\psi\bar{1}}$ are identified with each other and are projected to the $\epsilon$ anyon in the $\mathbb{Z}_2$ toric code layers. Therefore, after the fracton condendation, the extended ICN and the corresponding fraction phase becomes decouple layers of the $\mathbb{Z}_2$ toric code model and phases. 

Fracton condensation can arise in any extended cage-net models. For a simpler example, the fraction condensation in the XC model would result in the trivial phase. For fracton condensation in more complicated cases, we shall discuss in Section \ref{sec: conclusion and outlook}.

\section{From ICN to X-cube via non-Abelian p-loop condensation}
\label{non-Abelian p-loop cond}

In this section, we shall construct the non-Abelian p-loop condensation -- specifically $\sigma\bar\sigma$-loop condensation -- within the ICN model, such that it causes a phase transition from the ICN model to the X-cube model. We shall see that $\sigma\bar\sigma$-loops cannot condense alone but have to trigger the condensation of fracton $\mathfrak{f}_{\psi\bar{\psi}}$. To construct the local projectors $P_E$ that condense $\sigma\bar\sigma$-loops, we recall that anyon $\sigma\bar\sigma$ has a 2-dimensional internal space spanned by two dyonic states: $(\sigma\bar\sigma, 1)$ and $(\sigma\bar\sigma, \psi)$. As such, $\sigma\bar\sigma$ can condense in two possible ways (characterized by two Goldstone modes)\cite{zhao_landau-ginzburg_2025,zhao_nonabelian_2025}, i.e., condensing the pure chargeon state $(\sigma\bar\sigma,1)$ and condensing the state $(\sigma\bar\sigma,\psi)$. In this paper, we consider condensing the former only, namely $(\sigma\bar\sigma,1)$-loops.

There are two subsets of $L_\text{ICN}$ that are closed under the Ising fusion rules. One of them is the $L_{\mathrm{d}\mathbb{Z}_2}$ defined in \eqref{eq: LdZ2}, corresponding to the fracton condensation. Therefore, we should choose the other subset 
\begin{equation}
    L_{\text{XC}}=\{1|1,\psi|\psi\}
\end{equation}
as our target subset. We will show that condensing $\left(\sigma\Bar{\sigma},1\right)$-loops in the extended ICN phase projects out the principal edge values $\{1|\psi,\psi|1,\sigma|\sigma\}$ from $L_\text{ICN}$, resulting in a child model that is identical to the X-cube model defined on the truncated cubic lattice.

\subsection{$(\sigma\bar{\sigma},1)$-loop condensation and XC model}

The $(\sigma\Bar{\sigma},1)$-loop condensation operator $P^{\text{XC} \mid \text{ICN}}$ is formally written as
\begin{equation}
    P^{\text{XC} \mid \text{ICN}}= \prod_E P^{\text{XC} \mid \text{ICN}}_E,
\end{equation}
where
\begin{equation}
    P^{\text{XC} \mid \text{ICN}}_E=\pi_{1}\mathbf{1}_E+\pi_{L}L^{\sigma\bar{\sigma}}_E+\pi_{(C,1\psi)}C^{1\bar{1},\psi\bar{\psi}}_E+\pi_{(C,1\sigma)}C^{1\bar{1},\sigma\bar{\sigma}}_E + \pi_{(C,\psi\sigma)}C^{\psi\bar{\psi},\sigma\bar{\sigma}}_E.
    \label{sigma loop cond general}
\end{equation}
Here, the operators $\mathbf{1}_E, L^{\sigma\bar{\sigma}}_E, C^{1\bar{1},\psi\bar{\psi}}_E, C^{1\bar{1},\sigma\bar{\sigma}}_E$ and $C^{\psi\bar{\psi},\sigma\bar{\sigma}}_E$ are defined in Table \ref{tab: ICN p-loop cond terms}.
\begin{table}[!h]
    \centering
    \begin{tabular}{|c|c|}
        \hline 
        \textbf{Notation} & \textbf{Explicit Form} \\
        \hline 
        $\mathbf{1}_E$ & $\frac{1}{2}\left( W^{1\Bar{1} ;1,1}_E V^{1\Bar{1} ;1,1}_E +W^{\psi \Bar{\psi } ;1,1}_E V^{\psi \Bar{\psi } ;1,1}_E\right)$ \\
        \hline 
        $L^{\sigma \bar{\sigma}}_E$ & $W^{\sigma\bar{\sigma};1,1}_E V^{\sigma\bar{\sigma};1,1}_E$\\
        \hline 
        $C^{1\bar{1}, \psi \bar{\psi}}_E$ & $\frac{1}{2}\left( W^{1\Bar{1} ;1,1}_E V^{\psi \Bar{\psi } ;1,1}_E +W^{\psi \Bar{\psi } ;1,1}_E V^{1\Bar{1} ;1,1}_E \right)$\\
        \hline 
        $C^{1\bar{1},\sigma\bar{\sigma}}_E$ & $\frac{1}{2}\left( W^{1\Bar{1} ;1,1}_E V^{\sigma \Bar{\sigma } ;1,1}_E +W^{\sigma \Bar{\sigma } ;1,1}_E V^{1\Bar{1} ;1,1}_E \right)$ \\
        \hline 
        $C^{\psi\bar{\psi},\sigma\bar{\sigma}}_E$ & $\frac{1}{2}\left( W^{\psi\Bar{\psi} ;1,1}_E V^{\sigma \Bar{\sigma } ;1,1}_E +W^{\sigma \Bar{\sigma } ;1,1}_E V^{\psi\Bar{\psi} ;1,1}_E \right)$ \\
        \hline
    \end{tabular}
    \caption{Explicit forms of the five independent terms in \eqref{sigma loop cond general}.}
    \label{tab: ICN p-loop cond terms}
\end{table}

The coefficients $\pi$ in \eqref{sigma loop cond general} are determined by the projector condition
\begin{equation}
    P^{\text{XC} \mid \text{ICN}}_E \ket{\psi\varphi}=\delta_{{j_E| k_E}\in L_\text{XC}}\ket{\psi\varphi}.
    \label{sigma loop cond eq}
\end{equation}

This projector gaps out $\sigma | \sigma$ degree of freedom in every principal edge, effectively condenses $\psi\bar{\psi}$ anyon in every layer (2d HGW model) and causes the following mappings of dyon types in every layer:
\begin{equation}
\begin{split}
    & (1\bar{1},1), \quad (\psi\bar{\psi},1) \mapsto (1,1), \\
    & (1\bar{\psi},\psi), \quad (\psi\bar{1},\psi) \mapsto (\epsilon,\psi), \\
    & (\sigma\bar{\sigma},1) \mapsto (e,1), \\
    & (\sigma\bar{\sigma},\psi) \mapsto (m,\psi), \\
    & (1\bar{\sigma},\sigma), \quad (\psi\bar{\sigma},\sigma), \quad (\sigma\bar{1},\sigma), \quad (\sigma\bar{\psi},\sigma) \mapsto \emptyset.   
    \label{eq: dyon type mapping}
\end{split}
\end{equation}

Solving  \eqref{sigma loop cond eq} yields the explicit form of the $(\sigma\Bar{\sigma},1)$-loop condensation term:
\begin{equation}
    -\Lambda\sum_E P^{\text{XC} \mid \text{ICN}}_E=-\Lambda\sum_E \left(\frac{1}{4}\left(\mathbf{1}+C_{1\bar{1} , \psi \bar{\psi}}\right)+\frac{1}{2} L_{\sigma\bar{\sigma}}\right).
    \label{sigma loop cond term}
\end{equation}

This operator has a clear physical meaning: In the limit where the coupling $\Lambda$ in \eqref{dHGW Hamil with p-loop} approaches infinity, the ground states of the system (ICN + $(\sigma\bar\sigma,1)$-loop condensation) are the common $+1$ eigenstates of all local projectors $P^{\text{XC} \mid \text{ICN}}_E$. The cross term $C^{1\bar{1} , \psi \bar{\psi}}$ can suppress the additional fusion channels arising from the non-Abelian nature of $(\sigma\bar{\sigma},1)$, thereby ensuring that the $(\sigma\Bar{\sigma},1)$-loops fluctuate freely. Note that $C^{1\bar{1} , \psi \bar{\psi}}$ is exactly the fracton creation operator defined in \eqref{eq: fracton creation}, which means that during the $(\sigma\bar{\sigma},1)$-loop condensation, fracton $\mathfrak{f}_{\psi\bar{\psi}}$ must also be condensed. 

\paragraph{Child Hamiltonian}
The effective Hamiltonian of the child phase following $(\sigma\bar{\sigma},1)$-loop condensation is given by:
\begin{equation}
    H_\text{child}^\mathrm{eff} = -\sum_{(E,\mu)}\left(\sum_{P_d} Q_{P_d} - \sum_{P_o} Q_{P_o}^1\right) - \sum_C F_C=H_\text{XC}.
\end{equation}
Here, $Q_{P_d}$ represents the diamond-plaquette operator with the $\sigma$-component projected out (i.e., $Q_{P_d}$ excluding $Q_{P_d}^\sigma$), and the renormalized cube operator $F_C^{\text{XC}}$ takes the form:
\begin{equation}
    F_C = \prod_{P_o \in C} Q^{\psi}_{P_o},
\end{equation}
which is identical to the X-cube Hamiltonian we obtained in Section \ref{sec: xcube from dZ2}.

\subsection{Quasiparticle types in XC model: revisit}

In this section, we shall obtain quasiparticles in the XC model by projecting the quasiparticles in the ICN model via projector $P^{\text{XC} \mid \text{ICN}}$ \eqref{sigma loop cond eq}. As to be seen, the quasiparticle types in Eq. \eqref{ICN quasiparticle types} can be reproduced via the $(\sigma\bar\sigma,1)$-loop condensation in the ICN model. The two internal sectors of planon $\mathfrak{p}_{\sigma\Bar{\sigma}}$ in the ICN model, i.e., $\mathfrak{p}_{(\sigma\Bar{\sigma},1)}$ and $\mathfrak{p}_{(\sigma\Bar{\sigma},\psi)}$, are projected differently. The flux-free sector $\mathfrak{p}_{(\sigma\Bar{\sigma},1)}$ is projected to fracton $\mathfrak{f}_e$ in the XC model. For example, 
\begin{align}
\begin{split}
    & P^{\text{XC} \mid \text{ICN}} \left(\vcenter{\hbox{\includegraphics[height=12ex]{sigbarsig_vac.png}}} +\vcenter{\hbox{\includegraphics[height=12ex]{sigbarsig-psibarpsi.png}}} + \ldots \right) \\
    & =\vcenter{\hbox{\includegraphics[height=12ex]{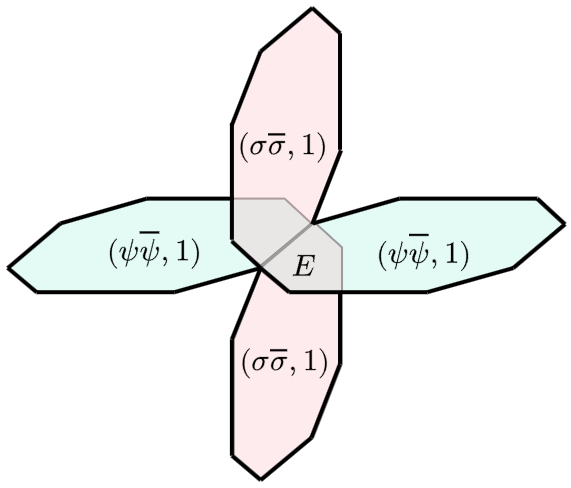}}} + \vcenter{\hbox{\includegraphics[height=12ex]{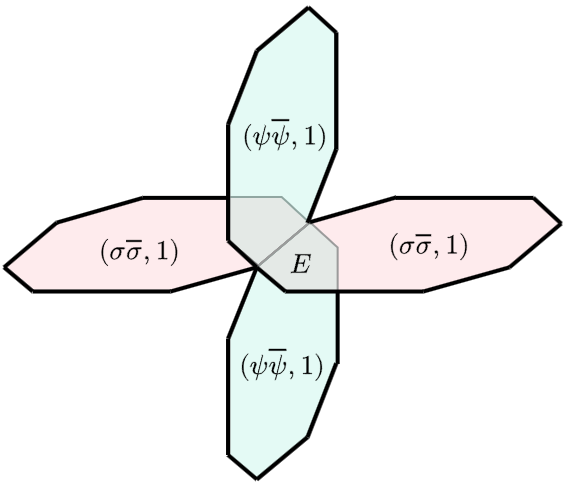}}} + \vcenter{\hbox{\includegraphics[height=12ex]{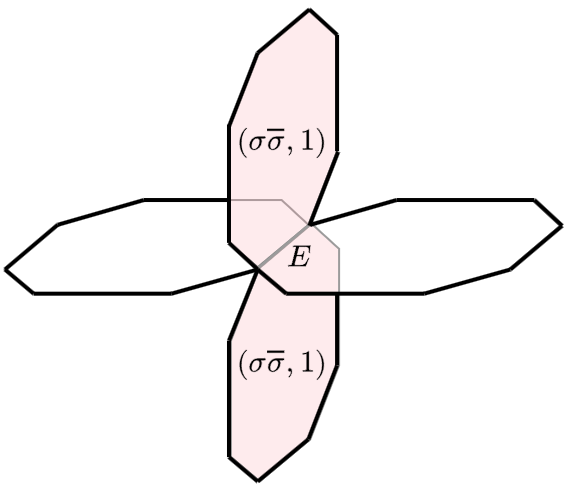}}} + \vcenter{\hbox{\includegraphics[height=12ex]{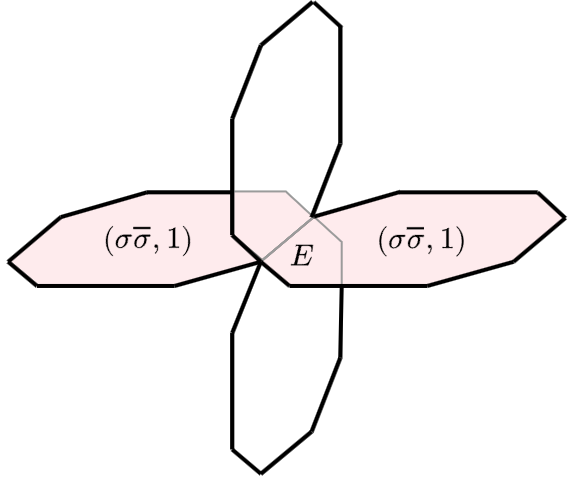}}} +\ldots\\
    &
    =\vcenter{\hbox{\includegraphics[height=12ex]{fracton_e_1.png}}} + \vcenter{\hbox{\includegraphics[height=12ex]{fracton_e_2.png}}} + \ldots,
\end{split}
\end{align}
which is precisely the state consisting of four fractons $\mathfrak{f}_e$ in the four cubes around principal edge $E$ in the XC model. The third row of the equation above follows the mappings in  \eqref{eq: dyon type mapping}.

The sector $\mathfrak{p}_{(\sigma\Bar{\sigma},\psi)}$ is 
 however projected out of the XC Hilbert space. For example,

 \begin{equation}
     P^{\text{XC} \mid \text{ICN}} \left(\vcenter{\hbox{\includegraphics[height=12ex]{sigbarsig,psi_vac.png}}} + \vcenter{\hbox{\includegraphics[height=12ex]{sigbarsig,psi-psibarpsi.png}}} + \ldots \right) = 0.
 \end{equation}
The different behavior of the two internal sectors of planon $\mathfrak{p}_{\sigma\bar{\sigma}}$ in the ICN model upon $(\sigma\bar{\sigma},1)$-loop condensation is a 3-dimensional counterpart of the splitting of anyons occuring in the scenario of anyon condensation in two-dimensional topological orders \cite{bais_condensate-induced_2009,hung_generalized_2015,hu_anyon_2022}. We manifest such a splitting at the level of states in our extended ICN model, as is done for anyon condensation in the HGW string-net model\cite{wang_electric-magnetic_2020,hu_electric-magnetic_2020,zhao_characteristic_2023}.

Any lineon or any planon with $\psi$ flux (such as $\mathfrak{p}_{1\bar{\psi}}$) is projected out of the XC Hilbert space.

An isolated star configuration of two perpendicular pairs of planons with a $\psi$ flux around a principal edge is projected to a state comprising two lineons in the XC model. For instance, the isolated star configuration of two perpendicular pairs of $\mathfrak{p}_{1\psi}$ and that of two perpendicular pairs of $\mathfrak{p}_{(\sigma\bar{\sigma},\psi)}$ are projected into the same state comprising two lineons $\mathfrak{l}_{mm}$ in the XC model:
\begin{equation}
\begin{split}
   & P^{\text{XC} \mid \text{ICN}} \left(\vcenter{\hbox{\includegraphics[height=12ex]{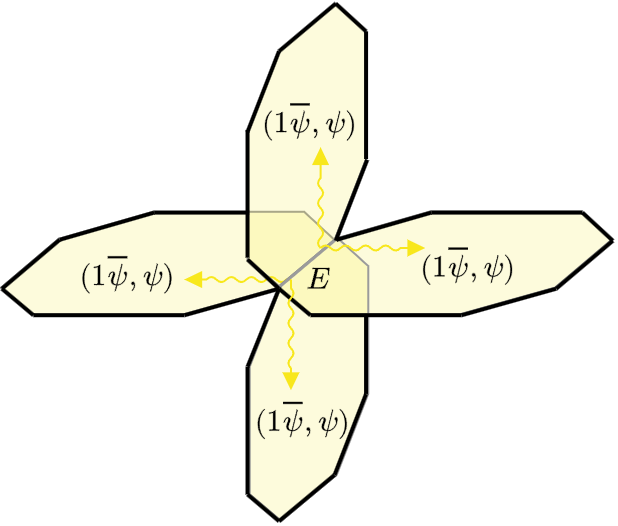}}} +\vcenter{\hbox{\includegraphics[height=12ex]{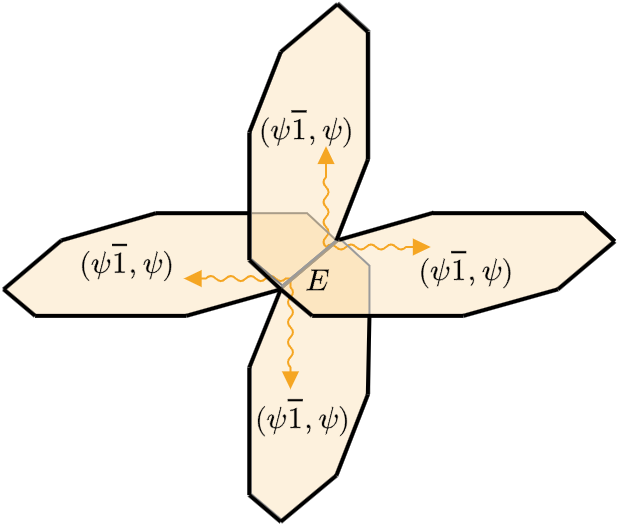}}} + \ldots \right)
    =P^{\text{XC} \mid \text{ICN}} \left(\vcenter{\hbox{\includegraphics[height=12ex]{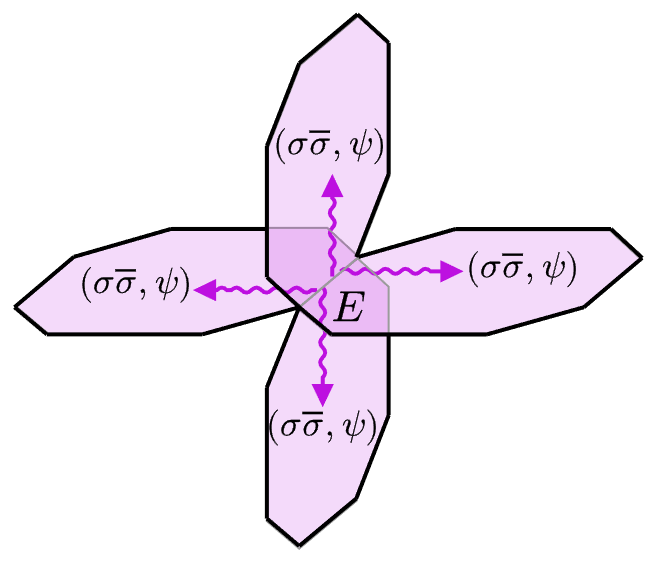}}} + \ldots\right) \\
    & =\vcenter{\hbox{\includegraphics[height=12ex]{inter_lineon_mm_1.png}}} + \vcenter{\hbox{\includegraphics[height=12ex]{inter_lineon_mm_2.png}}} + \vcenter{\hbox{\includegraphics[height=12ex]{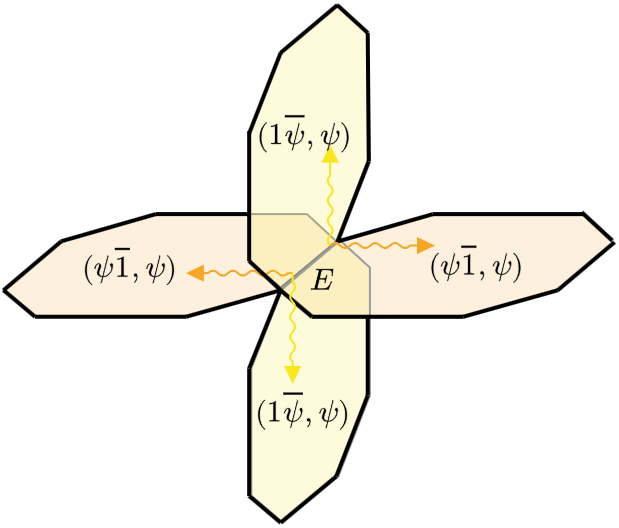}}} + \vcenter{\hbox{\includegraphics[height=12ex]{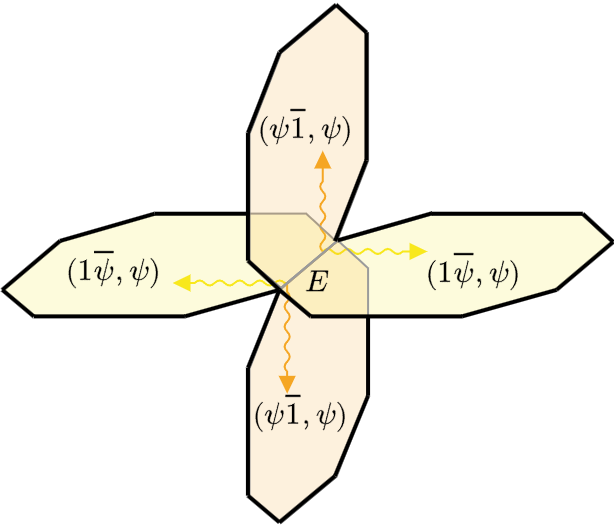}}} +\ldots\\
    &=\vcenter{\hbox{\includegraphics[height=12ex]{lineon_mm_1.png}}} + \vcenter{\hbox{\includegraphics[height=12ex]{lineon_mm_2.png}}} + \ldots,
\end{split}
\label{eq: lineon m m mapping}
\end{equation}

Similarly, we have the following projection:
\begin{align}
\begin{split}
    & P^{\text{XC} \mid \text{ICN}} \left(\vcenter{\hbox{\includegraphics[height=12ex]{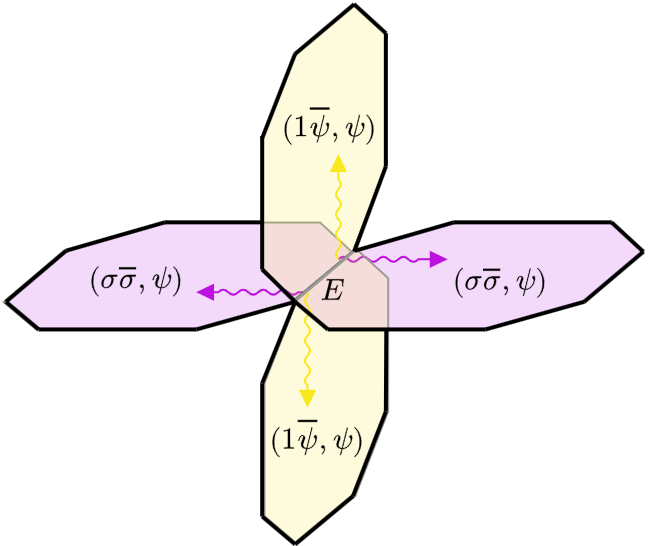}}} +\vcenter{\hbox{\includegraphics[height=12ex]{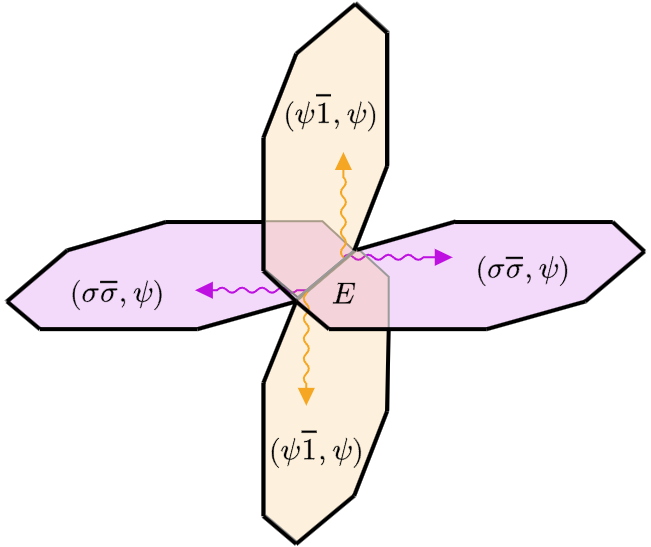}}} + \ldots \right) \\
    & =\vcenter{\hbox{\includegraphics[height=12ex]{inter_lineon_m_ep_1.png}}} + \vcenter{\hbox{\includegraphics[height=12ex]{inter_lineon_m_ep_2.png}}} + \vcenter{\hbox{\includegraphics[height=12ex]{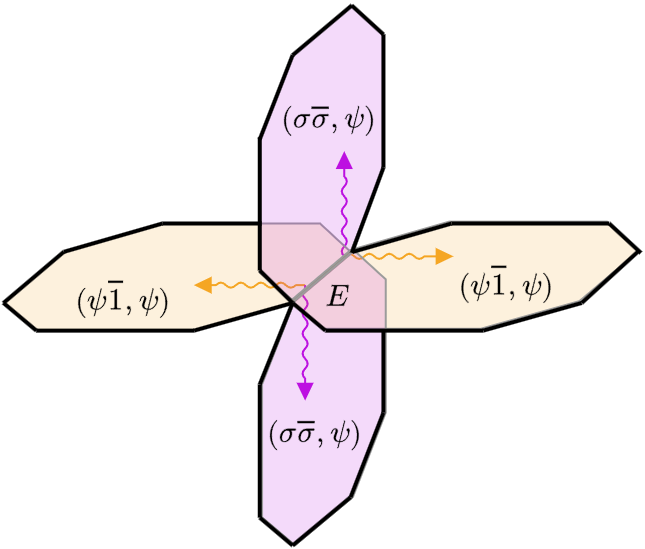}}} + \vcenter{\hbox{\includegraphics[height=12ex]{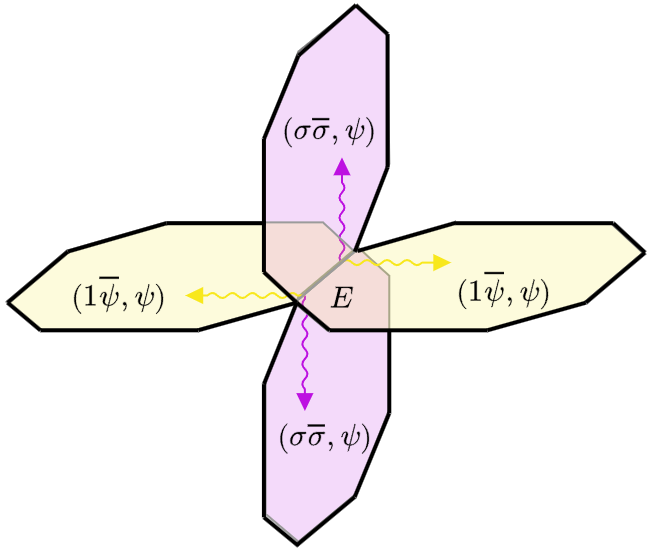}}} +\ldots\\
    &
    =\vcenter{\hbox{\includegraphics[height=12ex]{lineon_m_ep_1.png}}} + \vcenter{\hbox{\includegraphics[height=12ex]{lineon_m_ep_2.png}}} + \ldots,
\end{split}
\label{eq: lineon m epsilon mapping}
\end{align}
which is precisely the state consisting of two lineons $\mathfrak{l}_{m \epsilon}$ along the principal edge $E$ in the XC model. The third rows of the equations \eqref{eq: lineon m m mapping} and \eqref{eq: lineon m epsilon mapping} above follow the mappings in \eqref{eq: dyon type mapping}.

\section{Planon condensation}

Planon may also condense in the extended cage-net model. To be specific, we take the extended ICN model and condense planon $\mathfrak{p}_{(\sigma\bar{\sigma},1)}$ therein for example. The condensation operator of planon $\mathfrak{p}_{(\sigma\bar{\sigma},1)}$ is defined as:
\begin{equation}
    P^{\mathrm{Vac} \mid \text{ICN}}=\prod_E \left( P^{\mathrm{Vac} \mid \text{ICN}}_E \right), \quad P^{\mathrm{Vac} \mid \text{ICN}}_E = \frac{1}{8}\left(\mathbf{1}_E+C^{1\bar{1} , \psi \bar{\psi}}_E \right)+\frac{1}{2}C^{1\bar{1},\sigma\bar{\sigma}}_E+\frac{1}{4}L^{\sigma \bar{\sigma}}_E,
\end{equation}
where $\mathbf{1}_E, C^{1\bar{1} , \psi \bar{\psi}}_E, C^{1\bar{1},\sigma\bar{\sigma}}_E$ and $L^{\sigma \bar{\sigma}}_E$ are defined in Table \ref{tab: ICN p-loop cond terms}. 
The projector $P^{\mathrm{Vac} \mid \text{ICN}}$ gaps out all the nontrivial degrees of freedom in $L_\text{ICN}$:
\begin{equation}
    P^{\mathrm{Vac} \mid \text{ICN}}: L_\text{ICN}=\{1|1,\psi| \psi ,\sigma| \sigma ,1|\psi ,\psi| 1\} \longrightarrow L_\mathrm{Vac}=\{1|1\},
\end{equation}
leading a transition from the ICN model to a trivial model.

\section{Discussion and outlook}
\label{sec: conclusion and outlook}
In this work, we have addressed the challenge of non-Abelian p-loop and fracton condensation in the cage-net model by systematically extending the cage-net model. The original cage-net model was constructed by performing Abelian p-loop condensation in decoupled layers of the LW string-net model. Because the LW model lacks necessary degrees of freedom to capture its presumable full dyon spectrum, the original cage-net model was not able to classify its full quasiparticle (fractons, lineons, and planons) spectrum. We however construct the extended cage-net model by replacing the LW model with the HGW string-net model that contains the necessary tail degrees of freedom that resolve the full dyon spectrum of the string-net model. Therefore, the extended cage-net model we constructed permits a clear classification of its full quasiparticle spectrum. 

In our extended cage-net model
we defined p-loop condensation generally. This is benchmarked by rederiving the XC model from the $\mathrm{d}\mathbb{Z}_2$ model by condensing the $e$-loops in the latter. Then, we conducted the $\sigma\bar\sigma$-loop condensation in our extended ICN model, resulting in the child XC model. 

Remarkably, we found that the ICN fracton $\mathfrak{f}_{\psi\bar{\psi}}$ must condense along with the $\sigma\bar\sigma$-loops. We thus have an explicit definition of fracton condensation in our extended ICN and our extended cage-net model in general. We further found that the fracton condensation alone (without condensing the $\sigma\bar\sigma$-loops) in the extended ICN model would drive the model to be the $\mathrm{d}\mathbb{Z}_2$. In other words, the fracton condensation reduces and decouples the extended ICN model to be layers of two-dimensional toric code. It is possible to do fracton condensation in any extended cage-net model. This deserves future exploration. Our construction of condensation also extends to that of planons in the extended cage-net model. We explicitly showed as an example that condensing planon $\mathfrak{p}_{\sigma\bar{\sigma},1}$ in the extended ICN model results in the trivial phase.   
   
Since anyon $\sigma\bar\sigma$ has two internal unobservable sectors: the flux-free $(\sigma\bar\sigma,1)$ and dyonic $(\sigma\bar\sigma,\psi)$\cite{zhao_characteristic_2023}, a $\sigma\bar\sigma$-loop in the extended ICN model also has two internal sectors: $(\sigma\bar\sigma,1)$-loop and $(\sigma\bar\sigma,\psi)$-loop. These two sectors cannot condense simultaneously. In this paper, we only considered condensing the flux-free $(\sigma\bar\sigma,1)$-loops. Condensing the $(\sigma\bar\sigma,\psi)$-loops is much more involved and is left for future study.

We also witnessed that upon p-loop condensation, certain quasiparticles (such as planon $\mathfrak{p}_{\sigma\bar{\sigma}}$) in the extended ICN model can split, precisely analogous to the anyon splitting phenomemon in the scenario of anyon condensation in 2-dimensional topological orders. For example, upon $(\sigma\bar{\sigma},1)$-loop condensation in the extended ICN model, the two internal sectors of planon $\mathfrak{p}_{\sigma\bar{\sigma}}$ split: while one sector becomes fracton $\mathfrak{f}_e$, the other sector is confined in the resultant XC model.

While this work explicitly proves the transition between the ICN and XC models, we conjecture that non-Abelian p-loop condensation serves as a general mechanism for transitions between fracton phases. Future research may apply this framework to more complex unitary fusion categories to verify the universality of these mechanisms and to discover novel non-Abelian fracton orders.

\acknowledgments
YW is supported by NSFC Grant No.~12475001, the Shanghai Municipal Science and Technology Major Project (Grant No.~2019SHZDZX01), Science and Technology Commission of Shanghai Municipality (Grant No.~24LZ1400100), and the Innovation Program for Quantum Science and Technology (No.~2024ZD0300101). 
HS is supported by the National Natural Science Foundation of China (Grants No.~12522502, No.~12474145, and No.~12447101).
YW is grateful for the hospitality of the Perimeter Institute during his visit, where the main part of this work is done. This research was supported in part by the Perimeter Institute for Theoretical Physics. Research at Perimeter Institute is supported by the Government of Canada through the Department of Innovation, Science and Economic Development and by the Province of Ontario through the Ministry of Research, Innovation and Science. 

\appendix

\section{The HGW string-net model}
\label{HGW theory}

In this section we briefly review the HGW model defined in \cite{hu_full_2018,zhao_characteristic_2023}. The input data of the HGW model is a unitary fusion category $\mathcal{F}$, described by a finite set $L_\mathcal{F}$, whose elements are called \textit{simple objects}, equipped with three functions:
\begin{enumerate}
    \item $N : L_{\mathcal{F}}^3 \rightarrow \mathbb{N}$

    The function $N$ sets the fusion rules of the simple objects, satisfying

    \begin{equation}
        \sum_{e \in L_{\mathcal{F}}} N_{a b}^e N_{e c}^d=\sum_{e \in L_{\mathcal{F}}} N_{a e}^d N_{b c}^e, \quad N_{a b}^c=N_{c^* a}^{b^*}.
    \end{equation}

    There exists a special simple object $1 \in L_{\mathcal{F}}$, called the \textit{trivial obeject}, such that for $a,b \in L_{\mathcal{F}}$

    \begin{equation}
        N_{1 a}^b=N_{1 b}^a=\delta_{a b},
    \end{equation}

    where $\delta$ is the Kronecker symbol. For each $a \in L_{\mathcal{F}}$, there exists a unique unit simple object  $a^* \in L_{\mathcal{F}}$, called the $opposite object$ of $a$, such that

    \begin{equation}
        N_{a b}^1=N_{b a}^1=\delta_{b a^*}.
    \end{equation}

    We only consider the case where for any $a,b,c \in L_{\mathcal{F}}$, $N_{a b}^c=0$ or 1. In this case, we define

    \begin{equation}
        \delta_{a b c}=N_{a b}^{c^*} \in\{0,1\}.
    \end{equation}
    
    \item $d : L_\mathcal{F} \rightarrow \mathbb{R}^+$

    The function $d$ returns the $quantum  \ dimensions$ of the simple objects in $L_\mathcal{F}$. It is the largest eigenvalues of the fusion matrix and forms the 1-dimensional representation of the fusion rule.

    \begin{equation}
        d_a d_b=\sum_{c \in L_{\mathcal{F}}} N_{a b}^c d_c.
    \end{equation}

    In particular, $d_1=1$, and for any $a \in L_\mathcal{F}, d_a=d_{a^*} \geq 1$.
    
    \item $G : L_{\mathcal{F}}^6 \rightarrow \mathbb{C}$

    The function $G$ defines the \textit{6j-symbols} of the fusion algebra. It satisfies

    \begin{equation}
        \begin{gathered}
        \sum_n d_n G_{v^* u^* a}^{p q n} G_{j^* i^* b}^{u v n} G_{q^* p^* c}^{i j n}=G_{i^* p u^*}^{a b c} G_{v q^* j}^{c^* b^* a^*}, \quad \sum_n d_n G_{k l n}^{i j p} G_{l^* k^* n}^{j^* i^* q}=\frac{\delta_{p q^*}}{d_p} \delta_{i j p} \delta_{k l q}, \\
        G_{k l n}^{i j m}=G_{i j n^*}^{k l m^*}=G_{n k^* l^*}^{m i j}=\alpha_m \alpha_n {G_{l^* k^* n}^{j^*  i^* m^*}}, \quad \left|G_{1 b c}^{a b c}\right|=\frac{1}{\sqrt{d_b d_c}} \delta_{a b c},
        \end{gathered}
    \end{equation}

    where $\alpha_{a_m}=G_{1 m^* m}^{1 m m^*} \in \{\pm 1\}$ is the Frobenius-Schur indicator of simple object $m$.
    
\end{enumerate}

The Hamiltonian of the HGW model reads:
\begin{equation}
    H:=-\sum_{\text {Plaquettes } P} Q_P, \quad Q_P:=\frac{1}{D} \sum_{s \in L_{\mathcal{F}}} Q_P^s, \quad D:=\sum_{a \in L_{\mathcal{F}}} d_a^2,
\end{equation}

where operator $Q_P^s$ acts on edges surrounding  plaquette $P$ and has the following matrix elements on a hexagonal plaquette:
\begin{equation}
    Q_P^s\vcenter{\hbox{\includegraphics[height=12ex]{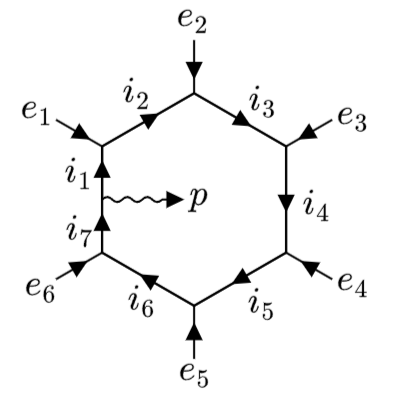}}}:=\delta_{p,0}\sum_{j_k \in L_\mathcal{F}}\prod_{k=1}^6(\left(\sqrt{d_{i_k} d_{j_k}} G_{s j_{k+1}^* j_k}^{e_k i_k i_{k+1}^*}\right))\vcenter{\hbox{\includegraphics[height=12ex]{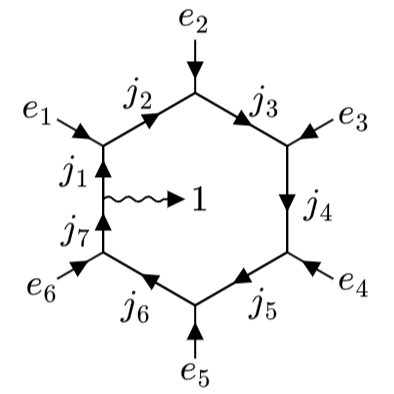}}}.
\end{equation}

Here we only show the actions of the $Q_P$ operators on a hexagonal plaquette. The matrix elements of $Q_P$ operators on other types of plaquettes, such as $Q_{P_o}$ for octagonal plaquettes (containing eight $6j$-symbols) and $Q_{P_d}$ for diamond-like plaquettes (containing four $6j$-symbols), are defined similarly. It turns out that $Q_P$ satisfies the following properties:
\begin{equation}
    \left(Q_P^s\right)^{\dagger}=Q_P^{s^*}, \quad Q_P^r Q_P^s=\sum_{t \in L_{\mathcal{F}}} N_{r s}^t Q_P^t, \quad Q_P^2=Q_P, \quad Q_{P_1} Q_{P_2}=Q_{P_2} Q_{P_1},
\end{equation}

which summand $Q_P$ in the Hamiltonian $H$ are commuting projectors, so the Hamiltonian is exactly solvable.

The fundamental setup of the string-net model is characterized by assigning a simple object from $L_{\mathcal{F}}$ to each edge and tail, while enforcing a vertex constraint: $\delta_{i j k}=1$ for the trio of incident edges or tails converging at a vertex, all directed toward the vertex and sequentially labeled counterclockwise by $i, j, k \in L_{\mathcal{F}}$. It is permissible to invert the direction of any edge or tail and simultaneously apply a conjugation to its label, expressed as $j \rightarrow j^*$, without altering the configuration. The Hilbert space $\mathcal{H}$ associated with the model is composed of the span of all conceivable configurations of these labels on the edges and tails.

The ground-state subspace $\mathcal{H}_0$ of the system is the projection:

\begin{equation}
    \mathcal{H}_0=\left[\prod_{\text {Plaquettes } P} Q_P\right] \mathcal{H} .
\end{equation}

\subsection{Topological Features}

In this appendix we briefly review the topological characteristics of the ground-state subspace of the HGW model. Lattices possessing identical topologies can be interconverted via \textit{Pachner moves} (collectively denoted as $\mathcal{T}$), corresponding to unitary linear mappings between the Hilbert spaces of two string-net models that utilize the same input fusion category on distinct lattices. The ground states exhibit invariance under these linear mappings. Three types of fundamental Pachner moves exist, with associated linear transformations specified as follows:
\begin{equation}
    \begin{gathered}
       \mathcal{T}\vcenter{\hbox{\includegraphics[height=10ex]{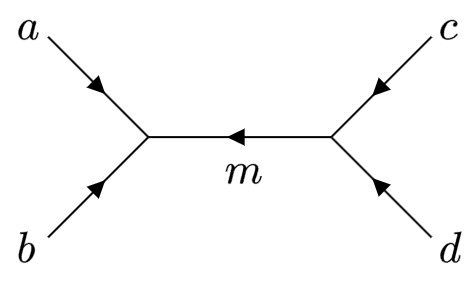}}}=\sum_{n \in L_{\mathcal{F}}} \sqrt{d_m d_n} G_{c d n}^{a b m}\vcenter{\hbox{\includegraphics[height=14ex]{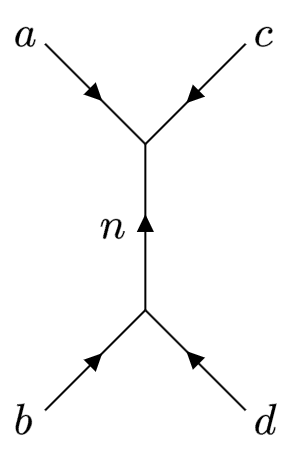}}}, \\
        \mathcal{T}\vcenter{\hbox{\includegraphics[height=12ex]{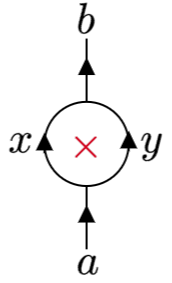}}}=\sqrt{\frac{d_x d_y}{d_i}}\delta_{i j}\delta_{x y i^*}\vcenter{\hbox{\includegraphics[height=12ex]{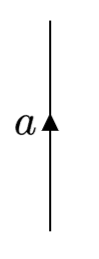}}}, \\
        \mathcal{T}\vcenter{\hbox{\includegraphics[height=12ex]{pacher_moves_4.png}}} = \frac{1}{D} \sum_{x y \in L_{\mathcal{F}}} \sqrt{\frac{d_x d_y}{d_i}} \delta_{x y i^*}\vcenter{\hbox{\includegraphics[height=12ex]{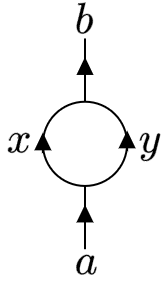}}}.
    \end{gathered}
\end{equation}

Here we use red $\textcolor{red}{\times}$ to mark the plaquettes to contract. Any other Pachner moves and their corresponding unitary transformations of Hilbert spaces are compositions of these three
elementary moves.

We also want to emphasize that different selections of the edge to which the tail is attached
are equivalent. The equivalence of states in such Hilbert
spaces is established by the following linear transformation $\mathcal{T}^\prime$:

\begin{equation}
    \mathcal{T}^\prime \vcenter{\hbox{\includegraphics[height=14ex]{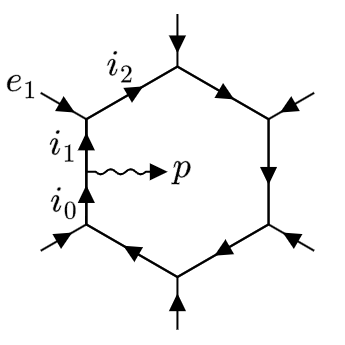}}}= \sum_{j \in L_{\mathcal{F}}} \sqrt{d_{i_1} d_j} G_{i_0 p^* j}^{i_2^* e_1 i_1} \vcenter{\hbox{\includegraphics[height=14ex]{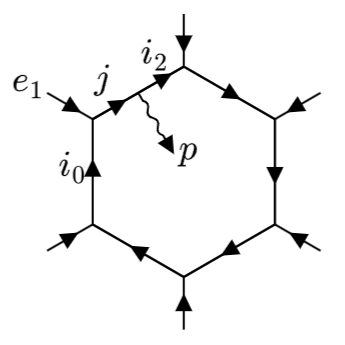}}}.
\end{equation}

The states where tails attach to other edges can be obtained recursively in this manner.

For convenience, in certain cases, we will temporarily incorporate auxiliary states with
multiple tails within a single plaquette. These states, despite having multiple tails in one
plaquette, are all equivalent to states within the Hilbert space:

\begin{equation}
    \vcenter{\hbox{\includegraphics[height=14ex]{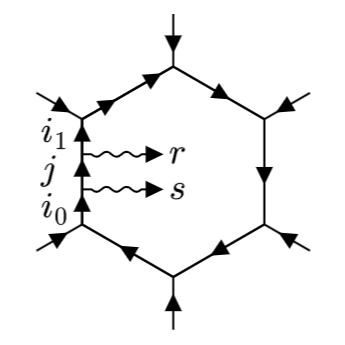}}}=\sum_{u \in L_{\mathcal{F}}} \sqrt{d_j d_p} G_{i_0 s^* p}^{r^* i_1^* j}\ \vcenter{\hbox{\includegraphics[height=14ex]{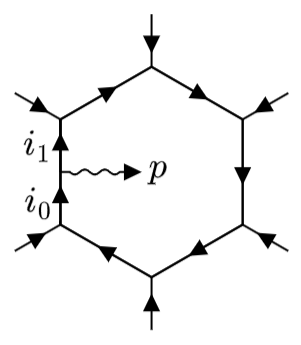}}}.
\end{equation}

\subsection{Excited States}

We start with the simplest excited states with a pair of anyons in two adjacent plaquettes with a common edge $E$. This state can be generated by ribbon operator $W_E^{J;p,q}$:
\begin{equation}
    W_E^{J;p,q}\vcenter{\hbox{\includegraphics[height=14ex]{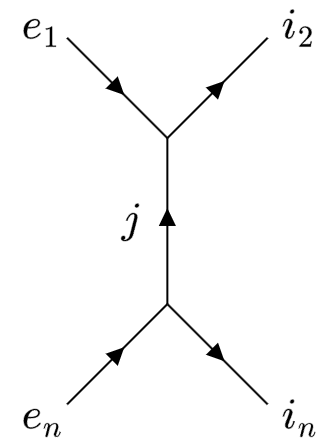}}}:=\sum_{k \in L_{\mathcal{F}}} \sqrt{\frac{d_k}{d_j}} \bar{z_{p q j}^{J ; k}} \vcenter{\hbox{\includegraphics[height=14ex]{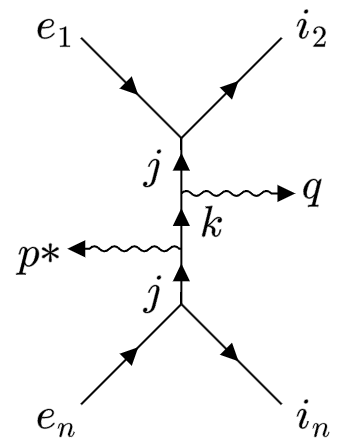}}},
    \label{rib op}
\end{equation}

where $j$ is the label on edge $E$, and $\Bar{z}$ is the complex conjugate. Here, $z_{p q j}^{J ; k}$ is called the \textit{half-braiding tensor} of anyon species $J$, defined by the following equation:

\begin{equation}
    \frac{\delta_{j t} N_{r s}^t}{d_t} z_{p q t}^{J ; w}=\sum_{u, l, v \in L_{\mathcal{F}}} z_{l q r}^{J ; v} z_{p l s}^{J ; u} \cdot d_u d_v G_{p^* w u^*}^{r^* s^* t} G_{q w^* v}^{s r j^*} G_{r v^* w}^{s^* u l^*}.
    \label{half-braid eq}
\end{equation}

The label $J$, called the \textit{anyon species}, labels different minimal solutions $z^J$ of \eqref{half-braid eq} that cannot be the sum of any other nonzero solutions. Categorically, anyon species $J$ are labeled by simple objects in the \textit{center} of UFC $\mathcal{F}$, a modular tensor category whose categorical data record all topological properties of the topological order that the string-net model describes, denoted as $\mathcal{Z}(\mathcal{F})$.

The statistics of anyon $J$ are recorded by its topological spin

\begin{equation}
    \theta_J=\frac{1}{d_p} \sum_{p \in L_{\mathcal{F}}} d_p z_{t t t}^{J ; p},
\end{equation}

where $t$ is an arbitrary charge of anyon $J$. The braiding of two anyons $J$ and $K$ is recorded by the modular $S$ matrix, whose matrix elements are

\begin{equation}
    S_{J K}=\sum_{p, q, k \in L_{\mathcal{F}}} d_k \bar{z}_{p q q}^{J ; k} \bar{z}_{q q p}^{K ; k}.
\end{equation}

An anyon $J$ has trivial self-statistics if $\theta_J=1$; two anyons $J$ and $K$ braid trivially if and only if $S_{J K}=d_J d_K$, where $d_J$ is the quantum dimension of anyon $J$, defined as

\begin{equation}
   d_J=\sum_{J^{\prime} \text { Charges } p} d_p . 
\end{equation}

\section{Categorical Data of Ising and $\text{Vec}(\mathbb{Z}_2)$ UFC}
\label{doubled ising data}

In this appendix we give the categorical data (including fusion rules, quantum dimensions, non-zero $6j$ symbols $G$, and nonzero components of half-braiding tensors $z$) of Ising and $\text{Vec}(\mathbb{Z}_2)$ UFC. 

The HGW model with input Ising UFC describes the doubled Ising topological phase. The simple objects of Ising UFC are $L=\{1,\psi,\sigma\}$.

\begin{enumerate}
    \item Quantum dimensions
        \begin{equation}
            d_1=d_\psi=1, \quad d_\sigma=\sqrt{2}.
        \end{equation}

    \item Fusion rules
        \begin{equation}
            \delta_{1 1 1}=\delta_{1 \psi \psi }=\delta_{1 \sigma \sigma}=\delta_{\psi \sigma \sigma}=1.
        \end{equation}

    \item Nonzero $6j$ symbols 

    The nonzero $6j$ symbols are solutions to the pentagon identity. Here we list some of the nonzero $6j$ symbols, while the other nonzero $6j$ symbols can be obtained through the tetrahedral symmetry.
    \begin{equation}
        \begin{aligned}
        & G_{111}^{111}=1, \quad G_{\sigma \sigma \sigma}^{111}=\frac{1}{\sqrt[4]{2}}, \quad G_{\psi \psi \psi}^{111}=1, \quad G_{1 \sigma \sigma}^{1 \sigma \sigma}=\frac{1}{\sqrt{2}}, \\
        & G_{\sigma \psi \psi}^{1 \sigma \sigma}=\frac{1}{\sqrt[4]{2}}, \quad G_{\psi \sigma \sigma}^{1 \sigma \sigma}=\frac{1}{\sqrt{2}}, \quad G_{1 \psi \psi}^{1 \psi \psi}=1, \quad G_{\sigma \psi \psi}^{\sigma \psi \psi}=-\frac{1}{\sqrt{2}} .
    \end{aligned}
    \end{equation}

    \item Nonzero components of $z$ tensors.

    Equation \eqref{half-braid eq} has 9 minimal solutions $z^{J_\text{DI}}$ with Ising input data, labeled by the 9 doubled-Ising anyon species. The nonzero components of these tensors are
    \begin{itemize}
    \item Anyon $1\bar{1}$ is the trivial anyon with trivial flux type $L_{1\bar{1}} = \{1\}$.
    \[
    [z_{11}^{1\bar{1}}]_1^1 = [z_{\psi\psi}^{1\bar{1}}]_1^1 = [z_{\sigma\sigma}^{1\bar{1}}]_1^1 = 1.
    \]

    \item Anyon $\psi\bar{\psi}$ has a unique trivial flux type $L_{\psi\bar{\psi}} = \{1\}$.
    \[
    [z_{11}^{\psi\bar{\psi}}]_1^1 = [z_{\psi\psi}^{\psi\bar{\psi}}]_1^1 = 1, \quad [z_{\sigma\sigma}^{\psi\bar{\psi}}]_1^1 = -1.
    \]

    \item Anyon $\psi\bar{1}$ and $1\bar{\psi}$ both have a unique nontrivial flux type $L_{\psi\bar{1}} = L_{1\bar{\psi}} = \{\psi\}$.
    \[
    [z_{1\psi}^{\psi\bar{1}}]_\psi^\psi = [z_{1\psi}^{1\bar{\psi}}]_\psi^\psi = 1, \quad [z_{\psi 1}^{\psi\bar{1}}]_\psi^\psi = [z_{\psi 1}^{1\bar{\psi}}]_\psi^\psi = -1, \quad [z_{\sigma\sigma}^{\psi\bar{1}}]_\psi^\psi = i, \quad [z_{\sigma\sigma}^{1\bar{\psi}}]_\psi^\psi = -i.
    \]

    \item Anyon $\sigma\bar{1}, \sigma\bar{\psi}, 1\bar{\sigma}$, and $\psi\bar{\sigma}$ all have a unique nontrivial flux type $\sigma$.
    \begin{gather*}
    [z_{1\sigma}^{\sigma\bar{1}}]_\sigma^\sigma = [z_{1\sigma}^{\sigma\bar{\psi}}]_\sigma^\sigma = [z_{1\sigma}^{1\bar{\sigma}}]_\sigma^\sigma = [z_{1\sigma}^{\psi\bar{\sigma}}]_\sigma^\sigma = 1, \\
    [z_{\sigma\psi}^{\sigma\bar{1}}]_\sigma^\sigma = [z_{\psi\sigma}^{\sigma\bar{\psi}}]_\sigma^\sigma = i, \quad [z_{\sigma 1}^{1\bar{\sigma}}]_\sigma^\sigma = [z_{\psi\sigma}^{\psi\bar{\sigma}}]_\sigma^\sigma = -i, \\
    [z_{\sigma 1}^{\sigma\bar{1}}]_\sigma^\sigma = e^{\frac{i\pi}{8}}, \quad [z_{\sigma\psi}^{1\bar{\sigma}}]_\sigma^\sigma = e^{\frac{3i\pi}{8}}, \quad [z_{\sigma\sigma}^{\sigma\bar{\psi}}]_\sigma^\sigma = e^{\frac{5i\pi}{8}}, \quad [z_{\psi 1}^{\psi\bar{\sigma}}]_\sigma^\sigma = e^{\frac{7i\pi}{8}}, \\
    [z_{1\sigma}^{1\bar{\sigma}}]_\sigma^\sigma = e^{-\frac{i\pi}{8}}, \quad [z_{\sigma\psi}^{\sigma\bar{1}}]_\sigma^\sigma = e^{-\frac{3i\pi}{8}}, \quad [z_{\sigma\sigma}^{\psi\bar{\sigma}}]_\sigma^\sigma = e^{-\frac{5i\pi}{8}}, \quad [z_{\sigma 1}^{\sigma\bar{\psi}}]_\sigma^\sigma = e^{-\frac{7i\pi}{8}}.
    \end{gather*}

    \item Anyon $\sigma\bar{\sigma}$ has two flux types 1 and $\psi$.
    \[
    [z_{11}^{\sigma\bar{\sigma}}]_1^1 = 1, \quad [z_{\psi\psi}^{\sigma\bar{\sigma}}]_1^1 = -1, \quad [z_{1\psi}^{\sigma\bar{\sigma}}]_\psi^\psi = [z_{\psi 1}^{\sigma\bar{\sigma}}]_\psi^\psi = 1, \quad [z_{\sigma\sigma}^{\sigma\bar{\sigma}}]_1^\psi = [z_{\sigma\sigma}^{\sigma\bar{\sigma}}]_\psi^1 = 1.
    \]
\end{itemize}

\end{enumerate}

The HGW model with input $\text{Vec}(\mathbb{Z}_2)$ UFC describes the $\mathbb{Z}_2$ toric code phase. Let $\mathbb{Z}_2 = \{1, \psi\}$ with fusion rules, quantum dimensions and $6j$ symbols given by
\[
\delta_{111} = \delta_{1\psi\psi} = 1, \quad d_1=d_\psi=1, \quad G_{cdn}^{abm} = \delta_{abm} \delta_{bcn} \delta_{cdm} \delta_{dan}.
\]

The toric code topological phase possesses four anyon types:

\begin{itemize}
    \item \textbf{The trivial anyon} $1$ with flux types $L_1 = \{1\}$:
    \[
    [z_{11}^{1}]_1^1 = [z_{\psi\psi}^{1}]_1^1 = 1.
    \]

    \item \textbf{The pure fluxon} $m$ with nontrivial flux type $L_m = \{\psi\}$, which has a unique dyonic sector $(m, \psi)$:
    \[
    [z_{1\psi}^m]_\psi^\psi = [z_{\psi 1}^m]_\psi^\psi = 1.
    \]

    \item \textbf{The pure chargeon} $e$ with flux type $L_e = \{1\}$, which has a unique dyonic sector $(e, 1)$:
    \[
    [z_{11}^e]_1^1 = 1, \quad [z_{\psi\psi}^e]_1^1 = -1.
    \]

    \item \textbf{The composite} $\epsilon (= e \times m)$, with flux type $L_\epsilon = \{\psi\}$, which has a unique dyonic sector $(\epsilon, \psi)$:
    \[
    [z_{1\psi}^\epsilon]_\psi^\psi = 1, \quad [z_{\psi 1}^\epsilon]_\psi^\psi = -1.
    \]
\end{itemize}

\section{Ribbon Algebra of Doubled Ising HGW Model}
\label{Ising ribbon algebra}

Consider acting a shortest ribbon operator $W_E^{J_1;p, q}$ on edge $E$, it will create two anyons $(J_1,p^*)$ and $(J_1,q)$ on the two adjacent plaquettes. Then we act another shortest ribbon operator $W_E^{J_2;r, s}$ on the same edge. The anyons $(J_2,r^*)$ and $(J_2,s)$ created by $W_E^{J_2;r, s}$ will fuse with the anyons $(J_1,p^*)$ and $(J_1,q)$, results in a superposition state. This process can be seen as the multiplication of ribbon operators:

\begin{equation}
    W_E^{J_2; r, s} W_E^{J_1; p, q}=\sum_{J,a,b} N_J^{a b} W_E^{J; a, b},
    \label{rib alg}
\end{equation}
where $N_J^{a b}$ is the multiplication coefficient.

If we use \eqref{rib op} and \eqref{half-braid eq} to expand \eqref{rib alg}, we obtain:

\begin{equation}
    \sum_k v_a v_b v_k^2 \bar{z_{r s k}^{J_2 ; l}} \ \bar{z_{p q j}^{J_1 ; k}} G_{l s b}^{q j k} G_{p j a}^{l r k}=\sum_J N_J^{a b} \bar{z_{a b j}^{J ; l}}, \quad \forall j, l .
    \label{rib alg eq}
\end{equation}

The doubled-Ising HGW model has twelve independent shortest ribbon operators, generating a 12-dimensional ribbon algebra, with 144 multiplications of operators. For example, when $(J_1;p,q) \in \lbrace (\psi\Bar{\psi};1,1),(\sigma\Bar{\sigma};1,1), (\sigma\Bar{\sigma};\psi,\psi) \rbrace $ and $(J_2;r,s)$ can take any possible value, the multiplications of operators are listed in Table \ref{tab:Ising rib alg}. The other multiplications of operators can be similarly derived using \eqref{rib alg eq}.

\begin{table}[!h]
    \centering
    \begin{tabular}{|c|c|c|c|}
        \hline
        \diagbox{$W_E^{J_2;r,s}$}{$W_E^{J_1;p,q}$} & $W_{E}^{\psi \bar{\psi} ;1,1}$ & $W_{E}^{\sigma \bar{\sigma} ;1,1}$ & $W_{E}^{\sigma \bar{\sigma} ;\psi ,\psi}$ \\
        \hline
        $W_{E}^{1\bar{1} ;1,1}$ & $W_{E}^{\psi \bar{\psi} ;1,1}$ & $W_{E}^{\sigma \bar{\sigma} ;1,1}$ & $W_{E}^{\sigma \bar{\sigma} ;\psi ,\psi}$ \\
        \hline
        $W_{E}^{\psi \bar{\psi} ;1,1}$ & $W_{E}^{1\bar{1} ;1,1}$ & $W_{E}^{\sigma \bar{\sigma} ;1,1}$ & $W_{E}^{\sigma \bar{\sigma} ;\psi ,\psi}$ \\
        \hline
        $W_{E}^{\psi \bar{1} ;\psi ,\psi}$ & $W_{E}^{1\bar{\psi} ;\psi ,\psi}$ & $W_{E}^{\sigma \bar{\sigma} ;\psi ,\psi}$ & $-W_{E}^{\sigma \bar{\sigma} ;1,1}$ \\
        \hline
        $W_{E}^{1\bar{\psi} ;\psi ,\psi}$ & $W_{E}^{\psi \bar{1} ;\psi ,\psi}$ & $W_{E}^{\sigma \bar{\sigma} ;\psi ,\psi}$ & $-W_{E}^{\sigma \bar{\sigma} ;1,1}$ \\
        \hline
        $W_{E}^{\sigma \bar{\sigma} ;1,1}$ & $W_{E}^{\sigma \bar{\sigma} ;1,1}$ & $(W_{E}^{1\bar{1} ;1,1} +W_{E}^{\psi \bar{\psi} ;1,1})/2$ & $-(W_{E}^{\psi \bar{1} ;\psi ,\psi} +W_{E}^{1\bar{\psi} ;\psi ,\psi})/2$ \\
        \hline
        $W_{E}^{\sigma \bar{\sigma} ;\psi ,\psi}$ & $W_{E}^{\sigma \bar{\sigma} ;\psi ,\psi}$ & $(W_{E}^{\psi \bar{1} ;\psi ,\psi} +W_{E}^{1\bar{\psi} ;\psi ,\psi})/2$ & $(W_{E}^{1\bar{1} ;1,1} +W_{E}^{\psi \bar{\psi} ;1,1})/2$ \\
        \hline
        $W_{E}^{\sigma \bar{\sigma} ;1,\psi}$ & $-W_{E}^{\sigma \bar{\sigma} ;1,\psi}$ & 0 & 0 \\
        \hline
        $W_{E}^{\sigma \bar{\sigma} ;\psi ,1}$ & $-W_{E}^{\sigma \bar{\sigma} ;\psi ,1}$ & 0 & 0 \\
        \hline
        $W_{E}^{\sigma \bar{1} ;\sigma ,\sigma}$ & $W_{E}^{\sigma \bar{\psi} ;\sigma ,\sigma}$ & $(W_{E}^{1\bar{\sigma} ;\sigma ,\sigma} +W_{E}^{\psi \bar{\sigma} ;\sigma ,\sigma})/2$ & $-i(W_{E}^{1\bar{\sigma} ;\sigma ,\sigma} +W_{E}^{\psi \bar{\sigma} ;\sigma ,\sigma})/2$ \\
        \hline
        $W_{E}^{\sigma \bar{\psi} ;\sigma ,\sigma}$ & $W_{E}^{\sigma \bar{1} ;\sigma ,\sigma}$ & $(W_{E}^{1\bar{\sigma} ;\sigma ,\sigma} +W_{E}^{\psi \bar{\sigma} ;\sigma ,\sigma})/2$ & $-i(W_{E}^{1\bar{\sigma} ;\sigma ,\sigma} +W_{E}^{\psi \bar{\sigma} ;\sigma ,\sigma})/2$ \\
        \hline
        $W_{E}^{1\bar{\sigma} ;\sigma ,\sigma}$ & $W_{E}^{\psi \bar{\sigma} ;\sigma ,\sigma}$ & $(W_{E}^{\sigma \bar{1} ;\sigma ,\sigma} +W_{E}^{\sigma \bar{\psi} ;\sigma ,\sigma})/2$ & $i(W_{E}^{\sigma \bar{1} ;\sigma ,\sigma} +W_{E}^{\sigma \bar{\psi} ;\sigma ,\sigma})/2$ \\
        \hline
        $W_{E}^{\psi \bar{\sigma} ;\sigma ,\sigma}$ & $W_{E}^{1\bar{\sigma} ;\sigma ,\sigma}$ & $(W_{E}^{\sigma \bar{1} ;\sigma ,\sigma} +W_{E}^{\sigma \bar{\psi} ;\sigma ,\sigma})/2$ & $i(W_{E}^{\sigma \bar{1} ;\sigma ,\sigma} +W_{E}^{\sigma \bar{\psi} ;\sigma ,\sigma})/2$ \\
        \hline
    \end{tabular}
    \caption{Part of the Doubled-Ising Ribbon Algebra.}
    \label{tab:Ising rib alg}
\end{table}




\bibliographystyle{JHEP} 
\bibliography{biblio}    

\end{document}